\documentclass[prx,aps, nobalancelastpage,superscriptaddress,nolongbibliography, onecolumn]{revtex4-2}

\usepackage{color,amsthm,amsmath,amsxtra,amsfonts,dsfont,graphicx,bm,amssymb,ulem}
\usepackage[colorlinks=true,linkcolor=blue, citecolor=blue, urlcolor=blue, bookmarks]{hyperref}
\usepackage{centernot}
\usepackage[dvipsnames]{xcolor}
\usepackage{graphicx}
\usepackage{tikz}
\usepackage{braket}
\usepackage{multirow}
\usepackage{comment}
\usepackage{makecell}

\newcommand{\be}{\begin{equation}}
	\newcommand{\ee}{\end{equation}}
\newcommand{\bea}{\begin{eqnarray}}
	\newcommand{\eea}{\end{eqnarray}} 
\newcommand{\bse}{\begin{subequations}}
	\newcommand{\ese}{\end{subequations}}

\theoremstyle{plain}

\newcommand{\tr}{\mathrm{Tr}}

\usepackage{pifont}
\usepackage{tikz}
\usetikzlibrary{decorations.pathreplacing,calligraphy,decorations.markings}

\begin{document}
	\date{\today}

	\newcommand{\bbra}[1]{\<\< #1 \right|\right.}
	\newcommand{\kket}[1]{\left.\left| #1 \>\>}
	\newcommand{\bbrakket}[1]{\< \Braket{#1} \>}
	\newcommand{\pll}{\parallel}
	\newcommand{\nn}{\nonumber}
	\newcommand{\transp}{\text{transp.}}
	\newcommand{\nor}{z_{J,H}}
	
	\newcommand{\hL}{\hat{L}}
	\newcommand{\hR}{\hat{R}}
	\newcommand{\hQ}{\hat{Q}}

\title{Entanglement Asymmetry in Random Quantum Automata}

\begin{abstract}
We investigate the subsystem entanglement asymmetry in random quantum automaton ensembles, which are generated by permuting the basis states in the Hilbert space and applying global phase shifts. We compute the ensemble average of the $U(1)$ subsystem asymmetry in different connectivity geometries, showing that the late-time limit of the ensemble associated to a 2-local circuit geometry coincides with the all-to-all ensemble average. By focusing on different subsystem sizes, we demonstrate that, similarly to Haar-random circuits, the system locally symmetrizes. However, in sharp contrast to the Haar-random setting, the scale at which symmetrization happens depends on the initial state, a phenomenon we associate with the interplay of conservation of the participation entropy and the uniform exploration of charge sectors. Additionally, we connect the growth of the subsystem asymmetry to the subsystem coherence and show that their growth is characterized by the same symmetrization scale.
\end{abstract}

\author{Olalla A. Castro-Alvaredo}
\affiliation{Department of Mathematics, City St George's, University of London, EC1V 0HB, UK}

\author{D\'avid Sz\'asz-Schagrin}
\affiliation{Dipartimento di Fisica e Astronomia, Universit\`a di Bologna and INFN, Sezione di Bologna, via Irnerio 46, 40126 Bologna, Italy}
\author{Michele Mazzoni}
\affiliation{Dipartimento di Fisica e Astronomia, Universit\`a di Bologna and INFN, Sezione di Bologna, via Irnerio 46, 40126 Bologna, Italy}

\maketitle
	 
\tableofcontents

\section{Introduction}
In the past decade, quantum circuits have emerged as an extremely powerful tool in the study of both integrable and chaotic quantum many-body dynamics \cite{hayden2007black,bertini2019exact,nahum2017quantum,zhou2019emergent, bertini2019entanglement, gopalakrishnan2019unitary, zhou2020entanglement, foligno2024quantum, piroli2020exact,klobas2021exact,bertini2024east,wang2024exact,hosur2016chaos,vonKeyserlingk2018operator,nahum2018operator, claeys2020maximum, bertini2020scrambling}. In particular, in the context of out-of-equilibrium dynamics of chaotic quantum many-body systems, random unitary circuits (RUCs) provide a simple, tractable model that, via a discretization of the time evolution, allows one to obtain exact results on the spread of correlations and entanglement growth \cite{fisher2023random, hunter2019unitary,dalzell2022random,harrow2023approximate,mittal2023local,schuster2024random,laracuente2024approximate,mele2024introduction}.
Although RUCs capture the typical behaviour of chaotic systems, circuits built from different classes of random gates can display richer dynamics. One such model that has gained considerable interest lately is the random quantum automaton circuit (QAC) \cite{iaconis2020measurement,iaconis2021quantum}. The QAC dynamics features very similar physics to RUCs when acting on a generic initial state, however, when restricted to appropriately chosen computational basis the dynamics is completely classical and there is no superposition creation. Generally, the QA gates consist of a random permutation of the computational basis, followed by a random phase transformation. The QAC dynamics is closely related to that of random permutation circuits \cite{Bertini2025Permutation}, although the presence of random phases make it more analytically tractable. Similar sets of quantum operations which act classically on the computational basis states were studied in integrable condensed matter models \cite{gopalakrishnan2018facilitated,Gopalakrishnan2018}.

Quantum automaton circuits, together with random permutation circuits for local dimension larger than $q=2$, display features of quantum chaotic systems \cite{Bertini2025Chaotic}, although they generally produce sparse wavefunctions when acting on highly localized states in the Hilbert space. One of the most striking differences between RUCs and QACs is that the latter preserve the participation entropy of the initial state in the computational basis. The participation entropy quantifies the extent to which the state is localized (or spread) in the Hilbert space, and the measure of this localization cannot change in time under QAC evolution. This has profound consequences on the entanglement growth in quantum automaton circuits, as well as in random permutation circuits, which were explored in Refs. \cite{szasz2026entanglement, Bertini2025Permutation}: most notably, the participation entropy of the initial state places an upper bound on the entanglement entropy that can be reached in the stationary state.

The remarkable features of QACs can be better appreciated in the context of quantum resource theories (QRTs) \cite{Chitambar2019resources}. In a QRT, one indirectly studies a quantity of interest, called a resource, by defining the free states for that resource, which do not display it, and the free operations, which cannot increase it. Resourceful states, on the other hand, can display a large amount of the resource. A classic example of a QRT is that of entanglement \cite{EntanglementQRT1996Bennet,MixedStateEntanglement1996Bennet,EntanglementQRT2009Horodecki,QuantifyingEntanglement1997Vedral}, where free states (i.e. zero entanglement states) are product states, and free operations are local operations and classical communications (LOCC). Any permutation of the states in a given basis followed by a phase transformation is a free operation for the QRT of coherence \cite{Baumgratz2014coherence,Streltsov2017coherence,Chitambar2019resources,Saxena2020coherence,aditya2026coherence}, in which free states correspond to density matrices with no off-diagonal elements in a specified basis. Indeed, an incoherent initial state will remain so under evolution with a quantum automaton circuit, since no coherent superposition of basis states can develop if the participation entropy is preserved during the evolution. The QRT of coherence has recently been explored in the context of ergodic quantum many-body dynamics and random quantum circuits \cite{turkeshi2024hilbert,Anticoncentration2025Lami,tirrito2025universal,Anticoncentration2025Magni,sauliere2025universality,tirrito2024anticoncentration,aditya2025growth,AnticoncentrationDoped2026Magni}.
    
In this work, we focus on $U(1)$ entanglement asymmetry, introduced two decades ago as a well-behaved quantifier of the quantum resource of asymmetry, or frameness \cite{vaccaro2008tradeoff, brs-07, gour2009measuring}. In this QRT, free states are those which commute with a certain symmetry (in our case, global $U(1)$ symmetry), and free operations can only produce symmetric states. Resourceful states, the ones which do not display the symmetry, can be employed in experiments and communication protocols involving exchanges between parties that do not have a shared reference frame. More recently, entanglement asymmetry was adopted in the context of quantum many-body systems as a probe of explicit or spontaneous symmetry breaking in a quantum many-body state \cite{marvian2014extending, ares2023entanglement}. Since the first works on entanglement asymmetry in quantum many-body systems, the interest in this quantity has consistently grown -particularly in out-of-equilibrium settings, where symmetry restoration can display a quantum Mpemba effect \cite{ares2023lack,rylands2024microscopic, murciano2024entanglement, chalas2024multiple, bertini2024dynamics, rylands2024dynamical, klobas2024asymmetry, caceffo2024entangled,foligno2024non, yamashika2024entanglement, yamashika2024quenching, ferro2024non,turkeshi2024quantum, liu2024symmetry,liu2024mpembaloc, banerjee2024asymmetry, klobas2024translation, summer2025resource,yu2025symmetry,aditya2025mpemba,aditya2026FragmentedMpemba2026,MpembaReview}. Many significant works on entanglement asymmetry have appeared in the context of random circuits and typical quantum many-body states \cite{capizzi2024universal, BlackHole, chen2024ph, Ares25, russotto25u1, Joshi2026}, conformal and quantum field theory \cite{casini2020entropic,casini2021entropic,magan2021,benedetti2024,capizzi2023entanglement, chen2024renyi, fossati2024entanglement, benini2025entanglementCFT, Kusuki2024, fossati2024, fujimura2025entanglement}, in relation with generalized symmetries \cite{benini2025entanglemenHigher, benini25cat, GattoLamas25} and with other QRTs \cite{ares2026asymmetryNonGauss}. Moreover, the out-of-equilibrium behaviour of $U(1)$ entanglement asymmetry has been recently probed in quantum simulations \cite{joshi2024observing,Xu2025Observation, Joshi2026}.

We study two ensembles associated to quantum automatons. First, we focus on the setting where some initial pure state is transformed via a global all-to-all gate consisting of a random permutation of the computational basis states followed by a global phase transformation. We call this the quantum automaton ensemble (QAE). Later, we build random circuits consisting of two-body QA gates, where each gate can act on any two qubits randomly, independently of their distance. These 2-local circuits have attracted considerable attention lately \cite{piroli2020random,Ares25, szasz2026entanglement}, for two reasons. On the one hand, two-local circuits are known to be fast scramblers \cite{piroli2020random}, meaning 
that they equilibriate on a timescale that is logarithmic in the system size. On the other hand, due to the relaxation of the locality constraint and the site-permutation invariance of the evolution operator, some quantities such as the Rényi-2 entanglement entropy admit a Markovian evolution that can be captured through a set of ordinary linear differential equations. Remarkably, the number of linear ODEs is polynomial in the system size.
    
\begin{figure}[h!]
	\centering
    \includegraphics[width = \linewidth]{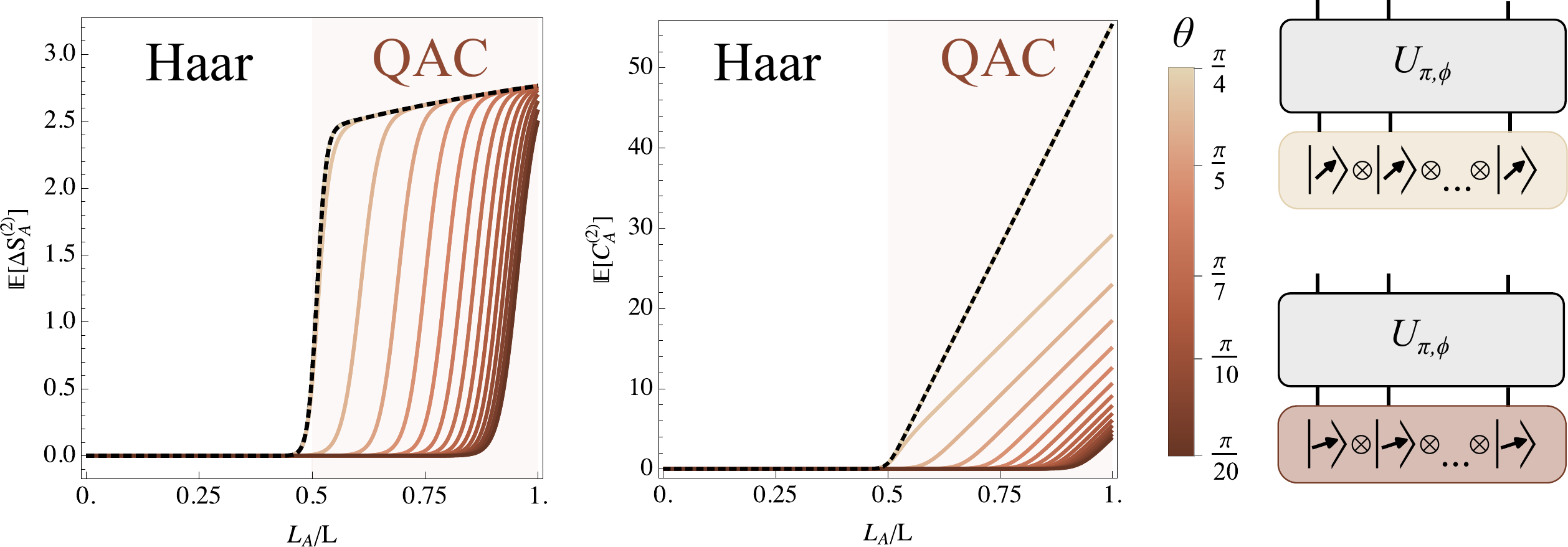}
	\caption{Schematic depiction of our main results. We apply a global random permutation of basis states followed by a random phase transformation to a homogeneous product state $\ket{\psi_0}=(\cos(\theta)|0\rangle+\sin(\theta)|1\rangle)^{\otimes L}$ (colored lines), with $L=80$. The curves for $\mathbb{E}[\Delta S_A^{(2)}]$ (left plot) are obtained using Eq. \eqref{eq:asym-global}, while those for $\mathbb{E}[C_A^{(2)}]$ (central plot) are obtained using the definition \eqref{eq:entropy:of_coherence_def} and the equations immediately preceding it. For $\theta = \pi/4$, the spread of the average R\'enyi-2 asymmetry and R\'enyi-2 subsystem coherence is akin to that of Haar random states (black dashed line), with both quantities being zero for subsystems with $L_A < L/2$. Decreasing $\theta$ results in a much more localized state in the Hilbert space and the onset of the averaged R\'enyi-2 asymmetry and R\'enyi-2 subsystem coherence cannot occur at $L_A=L/2$ because it is prevented by a too small participation entropy. The same is obtained if the global operation $U_{\pi,\phi}$ is replaced by the infinite-depth limit of a two-local circuit.}
	\label{fig:main_results}
\end{figure}

    
In both of these settings, we focus on $U(1)$ entanglement asymmetry averaged over the respective ensembles. In spin chains, it is naturally associated to a particle number symmetry or to the total magnetization. Remarkably, the stationary value of the $U(1)$ entanglement asymmetry of a subsystem in a RUC strongly depends on the size of the subsystem. In Ref. \cite{Ares25}, it was shown that an initial symmetric state will, at large times,  remain symmetric locally: the asymmetry of any subsystem smaller than half the size of the total system $L$ vanishes exponentially fast after a relaxation time. Conversely, the asymmetry of a subsystem larger than half the system size at large times reaches the typical value of half the maximum asymmetry (the origin of this typicality was explained in Ref. \cite{Mazzoni2025asymmetry}). The same picture holds for the average $U(1)$ asymmetry of a subsystem in a Haar random ensemble \cite{BlackHole}. As we will show in this work, in a quantum automaton ensemble and in quantum automaton circuits this picture changes drastically due to the conservation of the participation entropy (or total coherence) of the initial state. While on scales $l \leq L/2$ we always observe symmetrization, this scale depends on the initial state. Indeed, our first main result is that the competition between uniform exploration of charge sectors and preservation of the participation entropy leads to a shift of the threshold scale until which symmetrization occurs. In particular, this shift depends on the participation entropy of the initial state. This dependence is reminiscent of how the Page curve \cite{page1993average,page1993information}  for the average purity in QACs, as well as in random permutation circuits, depends on the initial state \cite{szasz2026entanglement}. However, the peculiar phenomenology of $U(1)$ entanglement asymmetry in QACs cannot be entirely inferred from the entanglement Page curves. Rather, as we show, the symmetrization scale is closely related to growth of
the average subsystem coherence. Indeed, our second main result is that the subsystem size at which late-time symmetrization ceases to occur is the same at which one starts to observe an extensive relative entropy of coherence in the subsystem. Figure \ref{fig:main_results} contains a sketch of our main results.

This work is organized in the following way. In Section \ref{sec:Preliminaries}, we define the quantum automaton ensemble and recall some technical notions on the computation of entanglement asymmetry in random ensembles. We relegate to Appendix \ref{sec:Choi representation for random quantum automata} a more detailed presentation of the replica vectorization formalism for random quantum automata. In Section \ref{sec:Averaged entanglement asymmetry in the quantum automaton ensemble}, we compute the average R\'enyi-2 asymmetry of homogeneous product states and Dicke states in the random automaton ensemble. In Section \ref{sec:Dynamics of entanglement asymmetry in a two-local quantum automaton circuit}, we derive a closed set of equations for the time evolution of the R\'enyi-2 asymmetries of all subsystems and we present numerical solutions. We leave to Appendix \ref{sec:Details on the dynamics in non-local QAC} the technical details of the derivation. In Section \ref{sec:Connection to subsystem coherence}, we present a decoupling inequality for the random automaton ensemble and explain the findings of the previous Sections by looking at the connection between subsystem asymmetry and subsystem coherence. We conclude in Section \ref{sec:Outlook and discussions}.

\section{Preliminaries}
\label{sec:Preliminaries}
In this section, we review some important notions concerning the quantum automaton ensemble and the $U(1)$ entanglement asymmetry. 

\subsection{The quantum automaton ensemble}
Throughout this paper, we consider a bipartite system $S=A \cup B$ with Hilbert space $\mathcal{H}=\mathcal{H}_A\otimes \mathcal{H}_B$. The system consists of $L$ qubits in a lattice of arbitrary spatial dimension, thus the local Hilbert space is $\mathcal{H}_k \simeq \mathbb{C}^2$, $k=1,\dots,L$, and:
\begin{equation}
\label{def:bipartite_Hilbert_space}
{\rm dim}\mathcal{H}_{A} = 2^{L_A} =: D_A, \quad {\rm dim}\mathcal{H}_{B} = 2^{L_B} =: D_B,
\end{equation}
with $L_A=|A|$, $L_B=|B|$, $L=L_A+L_B$ and $D:={\rm dim}\mathcal{H}=D_AD_B$. 

The quantum automaton ensemble (QAE), denoted by $\mathcal{E}_{\rm QAE}$, \cite{iaconis2021quantum,iaconis2020measurement} is the ensemble of operations $U$ with specified action on a basis of $\mathcal{H}$, which we choose to be the computational basis $\{\ket{s},\, s=0,\dots,D-1\}$. Namely:
\begin{equation}
\label{eq:random_perm_phase_definition}
 U\ket{s} = e^{i \phi_s}\ket{\pi(s)},
\end{equation}
where $\pi \in \mathcal{S}_D$ is an element of the permutation group on $D$ elements and $\phi_s \in [0,2\pi)$. The measure on the ensemble is the composition of the flat measure on the discrete permutation group and the uniform measure over $D$ copies of $U(1)$. In other words, this ensemble consists of all random (and uniformly chosen) permutations of the basis states, followed by independent and random phase transformations of the basis states\footnote{This ensemble was denoted by $\mathcal{E}_{\rm RPP}$ in the Supplemental material of \cite{szasz2026entanglement}}. In Section \ref{sec:Dynamics of entanglement asymmetry in a two-local quantum automaton circuit}, we will consider the ensemble of random quantum automaton circuits (QACs). For a fixed circuit depth $d$, we denote the corresponding ensemble by $\mathcal{E}_{\rm QAC}(d)$. We will focus on a circuit with non-local connectivity, where at each discrete time step two random qubits are coupled by a gate of the form \eqref{eq:random_perm_phase_definition}. By taking a suitable continuous time limit of the dynamics induced by a QAC, as explained in Section \ref{sec:Dynamics of entanglement asymmetry in a two-local quantum automaton circuit}, the ensemble $\mathcal{E}_{\rm QAC}(t)$ is well-defined, and it was shown in \cite{szasz2026entanglement} that, as $t\to \infty$ and the ensemble averages over $\mathcal{E}_{\rm QAC}(t)$ approach their stationary values, the R\'enyi-2 bipartite entanglement entropies averaged over the ensembles $\mathcal{E}_{\rm QAC}(\infty)$ and $\mathcal{E}_{\rm QAE}$ coincide. In Section \ref{sec:Dynamics of entanglement asymmetry in a two-local quantum automaton circuit}, we will show that this property holds also for the averaged R\'enyi-2 bipartite entanglement asymmetry. 

Any operation of the form \eqref{eq:random_perm_phase_definition} on a state $\ket{\psi}$ preserves its participation entropy \cite{stephan2009renyi,stephan2009shannon,alcaraz2013,stephan2014renyi,luitz2014participation}. As stated in the introduction, the participation entropy of a state measures its degree of localization with respect to a given basis of the Hilbert space. Namely, we define the R\'enyi-$n$ participation entropy of $\ket{\psi}$ as:
\begin{equation}
S_n^{\rm PE}(\ket{\psi}):= \frac{\log_2 I_n (\ket{\psi})}{1-n}\,,
\end{equation}
where we introduced the inverse participation ratio of the state \cite{kramer1993localization}:
\begin{equation}
\label{eq:second_inverse_participation_ratio}
I_n(\ket{\psi}):= \sum_s |\braket{s|\psi}|^{2n}\,, \quad 0 < I_n(\ket{\psi}) \le 1\,.
\end{equation} 
To see that the operation \eqref{eq:random_perm_phase_definition} preserves $I_n(\ket{\psi})$, it is sufficient to write $\ket{\psi}=\sum_s c_s\ket{s}$ and compute:
\begin{equation}
I_n(U\ket{\psi}) = \sum_s |\bra{s}U\ket{\psi}|^{2n} = \sum_s |c_{\pi^{-1}(s)}e^{i\phi_{\pi^{-1}(s)}}|^{2n} = \sum_s |c_{\pi^{-1}(s)}|^{2n} = \sum_s |c_s|^{2n}=I_n(\ket{\psi})\,,
\end{equation}
where we used the fact that each permutation $\pi$ is invertible. In the rest of the paper, the R\'enyi-2 participation entropy $S_2^{\rm PE}(\ket{\psi}):= -\log_2 I_2 (\ket{\psi})$ will play a crucial role.
    
Let us now express the average subsystem purity $\mathbb{E}[{\rm Tr}\rho_A^2]$ of a state $\rho_A={\tr}_{\mathcal{H}_B} \rho$ in the ensemble $\mathcal{E}_{\rm QAE}$. To do so, we employ the Choi-Jamio\l kowski (CJ) representation, or vectorization formalism \cite{watrous2018theory}. We review the essential results below, leaving the derivations and further details to Appendix \ref{sec:Choi representation for random quantum automata}. First, for every site $k \in A\cup B$ we define the following states in the replicated local Hilbert space $\mathcal{H}_k^{\otimes 4}$ \footnote{When the dimension of the local Hilbert space is $q\ge 2$, the indices in the sum run from $0$ to $q-1$}:
\begin{equation}
\label{eq:CJ_vectors}
\ket{I^+}_k := \sum_{a,b=0}^1 \ket{aa bb}_k, \quad \ket{I^-}_k := \sum_{a,b=0}^1 \ket{ab ba}_k, \quad \quad \ket{I^0}_k := \sum_{a=0}^1 \ket{aaaa}_k\,, 
\end{equation}
as well as $\ket{\mathcal{I}^{+}}_{A/B}= \otimes_{k \in A/B}\ket{I^+}_k$, $\ket{\mathcal{I}^{+}}= \otimes_{k \in A \cup B}\ket{I^+}_k$ and we analogously define $\ket{\mathcal{I}^-}$, $\ket{\mathcal{I}^0}$. The above states are not normalized, as indeed when the dimension of the local Hilbert space is $q=2$, the overlaps are:
\begin{equation}
{}_j\braket{I^+|I^+}_k = {}_j\braket{I^-|I^-}_k=4\delta_{j,k}\,, \quad  {}_j\braket{I^+|I^-}_k = {}_j\braket{I^-|I^0}_k={}_j\braket{I^+|I^0}_k={}_j\braket{I^0|I^0}_k=2\delta_{j,k}\,.
\end{equation} 
In the CJ representation, the 2-replica operator $\rho \otimes \rho$ is mapped to the state:
\begin{equation}
\label{eq:CJ_mapping_main}
\ket{\rho \otimes \rho} := \mathds{1} \otimes \rho(t) \otimes \mathds{1} \otimes \rho(t) \ket{\mathcal{I}^+} \in \mathcal{H}^{\otimes 4}\,.
\end{equation}

Let now $\rho_0=\ket{\psi_0}\bra{\psi_0}$ and $\rho=U\rho_0U^\dagger$, with the unitary matrix $U \in U(D)$ drawn from a certain ensemble $\mathcal{E}$, e.g. the Haar random ensemble or the random QAE. With the above definitions, the purity of the state $\rho_A$ averaged over $\mathcal{E}$ is expressed as follows:
\begin{equation}
\label{eq:average_purity_CJ_general}
\mathbb{E} [{\rm Tr} \rho_A^2] = {}_A\bra{\mathcal{I}^-}{}_B\bra{\mathcal{I}^+}\mathbb{E}[U^*\otimes U\otimes U^*\otimes U]\ket{\rho_0 \otimes \rho_0}\,.
\end{equation}
To compute the average purity, one needs the expression of the expectation value $\mathbb{E}[U^*\otimes U\otimes U^*\otimes U]$, which depends on the ensemble $\mathcal{E}$:
\begin{enumerate}
    \item For the Haar random ensemble:
    \begin{equation}
    \mathbb{E}_{\rm Haar}[U^*\otimes U\otimes U^*\otimes U]= \frac{1}{D^2-1}\left[\ket{\mathcal{I}^+}\bra{\mathcal{I}^+} + \ket{\mathcal{I}^-}\bra{\mathcal{I}^-}-\frac{1}{D}(\ket{\mathcal{I}^+}\bra{\mathcal{I}^-} + \ket{\mathcal{I}^-}\bra{\mathcal{I}^+})\right]\,,
    \end{equation}
    \item For $\mathcal{E}_{\rm QAE}$:
    \begin{align}
    \label{eq:second_moment_QAE_main}
    &\mathbb{E}_{\rm QAE}[U^*\otimes U\otimes U^*\otimes U] \nonumber \\= &\frac{1}{D(D-1)}\left[\ket{\mathcal{I}^+}\bra{\mathcal{I}^+} + \ket{\mathcal{I}^-}\bra{\mathcal{I}^-} +(D+1)\ket{\mathcal{I}^0}\bra{\mathcal{I}^0}-(\ket{\mathcal{I}^0}\bra{\mathcal{I}^-} + \ket{\mathcal{I}^-}\bra{\mathcal{I}^0} +\ket{\mathcal{I}^0}\bra{\mathcal{I}^+} + \ket{\mathcal{I}^+}\bra{\mathcal{I}^0})\right]\,.
    \end{align}
\end{enumerate}
Furthermore, it holds that:
\begin{equation}
\label{eq:initial_state_overlaps}
 \braket{{\mathcal{I}^-}|\rho_0 \otimes\rho_0} = \braket{{\mathcal{I}^+}|\rho_0 \otimes\rho_0}=1\,, \quad  
   \braket{{\mathcal{I}^0}|\rho_0 \otimes\rho_0} = I_2(\ket{\psi_0})\,.
\end{equation}
From the above, it follows that if $\mathcal{E}$ is the Haar random ensemble then:
\begin{equation}
\label{eq:Haar_random_average_purity}
\mathbb{E}_{\rm Haar}[{\rm Tr} \rho_A^2] = \frac{D_A + D_B}{D + 1}\,,
\end{equation}
which is a standard result that holds also in any 2-design ensemble \cite{nahum2018operator,mele2024introduction}. If instead $\mathcal{E}=\mathcal{E}_{\rm QAE}$, then \cite{szasz2026entanglement}:
\begin{equation}
\label{eq:QAE_averaged_purity}
\mathbb{E}_{\rm QAE}[{\rm Tr}\rho_A^2] = \frac{D_A + D_B -2 + I_2(\ket{\psi_0})[D-D_A-D_B+1]}{D-1}\,.
\end{equation}
Remarkably, unlike the Haar random expression Eq. \eqref{eq:Haar_random_average_purity}, the subsystem purity averaged over $\mathcal{E}_{\rm QAE}$ (\ref{eq:QAE_averaged_purity}) explicitly depends on the inverse participation ratio of the state $\ket{\psi_0}$, hence on some details of the initial state.  

\subsection{Entanglement asymmetry and charged partition functions}

 The $U(1)$ R\'enyi-$n$ entanglement asymmetry of the state $\rho_A$ is defined as the relative entropy \cite{ares2023entanglement}:
\begin{equation}
\Delta S_A^{(n)}:= S_n(\rho_{A,Q}) - S_n(\rho_A)\,,
\end{equation}
where the $U(1)$ charge $Q=\sum_{k=1}^L \ket{1}_k\bra{1}_k$ counts the number of qubits in the local state $\ket{1}$, the symmetrized density matrix is $\rho_{A,Q}= \sum_q \Pi_q^{(A)}\rho_A \Pi_q^{(A)}$, with $\Pi_q^{(A)}$ being the projectors onto the charge eigenspaces of $\mathcal{H}_A$, and $S_n(\rho)=(1-n)^{-1}\log {\rm Tr}\rho^n$. As above, we consider a state $\rho=U\rho_0 U^\dagger$ and we are interested in the expectation value of $\Delta S_A^{(n)}$, averaged over an ensemble $\mathcal{E}$:
\begin{equation}
\mathbb{E}[\Delta S_A^{(n)}]= \frac{\mathbb{E}[\log{\rm Tr}\rho_{A,Q}^n]}{1-n} - \frac{\mathbb{E}[\log{\rm Tr}\rho_{A}^n]}{1-n} \simeq \frac{\log\mathbb{E}[{\rm Tr}\rho_{A,Q}^n]}{1-n} - \frac{\log\mathbb{E}[{\rm Tr}\rho_{A}^n]}{1-n}\,,
\end{equation}
where the approximate equality holds as long as there is a self-averaging property in the ensemble (which is true for the quantities of interest both in the Haar random ensemble and in the QAE). The expectation values $\mathbb{E}{\rm Tr}\rho_{A,Q}^n$ can be computed via the charged partition functions:
\begin{equation}
\mathbb{E}[{\rm Tr}\rho_{A,Q}^n] = \int_{-\pi}^{\pi}\frac{d\alpha_1\dots d\alpha_n}{(2\pi)^n} \mathbb{E}[Z_A^{(n)}(\alpha_1,\dots,\alpha_n)]\,, \quad Z_A^{(n)}(\alpha_1,\dots,\alpha_n) = {\rm Tr} \prod_{j=1}^n \left[e^{i(\alpha_j-\alpha_{j+1})Q_A}\rho_A\right]\,,
\end{equation}
with $\alpha_{n+1}\equiv \alpha_1$ 
and $Q_A$ the $U(1)$ charge, restricted to subsystem $A$.

We are interested in the two-replica case, $n=2$. Here, setting $\alpha_1-\alpha_2:=\alpha$, one has $Z_A^{(2)}(\alpha_1, \alpha_2)=Z_A^{(2)}(\alpha)= {\rm Tr}[e^{i\alpha Q_A}\rho_A e^{-i\alpha Q_A} \rho_A]$. In the CJ representation, the average of this quantity reads:
\begin{equation}
\label{eq:charged_partition_function_CJ}
\mathbb{E}[ Z_A^{(2)}(\alpha)] = {}_A\bra{\mathcal{I}^-_\alpha}{}_B\bra{\mathcal{I}^+}\mathbb{E}[U^*\otimes U\otimes U^*\otimes U]\ket{\rho_0 \otimes \rho_0}\,,
\end{equation}
where we defined the state \cite{BlackHole, Ares25} $\ket{\mathcal{I}^-_\alpha}_A = \otimes_{k \in A}\ket{I^-_\alpha}_k$, $\ket{I^-_\alpha}_k = \sum_{a,b=0}^1 e^{i\alpha(a-b)}\ket{ab ba}_k$. Physically, the state $\ket{\mathcal{I}^-_\alpha}_A$ implements the insertion of the charge operators $e^{\pm i\alpha Q_A}$ between two copies of $\rho_A$. The overlaps of $\ket{I^-_\alpha}_k$ with the vectors in Eq. \eqref{eq:CJ_vectors} are:
\begin{equation}
\label{eq:I_alpha_overlaps}
 {}_j\braket{I^-_\alpha|I^0}_k={}_j\braket{I^-_\alpha|I^+}_k=2\delta_{j,k}\,, \quad {}_j\braket{I^-_\alpha|I^-}_k = 4f(\alpha) \delta_{j,k}\,, \quad f(\alpha):=\cos^2\left(\frac{\alpha}{2}\right)\,.
\end{equation}

\section{Averaged entanglement asymmetry in the quantum automaton ensemble}
\label{sec:Averaged entanglement asymmetry in the quantum automaton ensemble}

In this Section, we compute the averaged entanglement asymmetry $\mathbb{E}[\Delta S_A^{(2)}]$ in the ensemble $\mathcal{E}_{\rm QAE}$. As anticipated in the Introduction, our results are markedly different from those found in the Haar random ensemble \cite{BlackHole}. We consider two classes of states, the homogeneous product states and the Dicke states, characterized by a different scaling of the participation entropy $S_2^{\rm PE}(\ket{\psi})$ with the system size $L$, namely linear for the product states and logarithmic for the Dicke states. We show analytically how the profile of $\mathbb{E}[\Delta S_A^{(2)}]$, as a function of the subsystem size $L_A$, strongly depends on the participation entropy.

\subsection{Computation of the asymmetry}
We start by computing the expectation value of the charged partition function $Z_A^{(2)}(\alpha)$ in the QAE, for a state $\rho=U\ket{\psi_0}\bra{\psi_0}U^\dagger$. From the expression Eq. \eqref{eq:charged_partition_function_CJ}, together with the overlaps Eqs. \eqref{eq:initial_state_overlaps}, \eqref{eq:I_alpha_overlaps} one immediately obtains
\begin{equation}
\label{eq:averaged_charged_moment_QAE}
\mathbb{E}[Z_A^{(2)}(\alpha)] = \frac{D_A f(\alpha)^{L_A} + D_B -2 + I_2(\ket{\psi_0})[D-D_Af(\alpha)^{L_A}-D_B+1]}{D-1}\,.
\end{equation}
Notice that $Z_A^{(2)}(0)={\rm Tr}\rho_A^2$, thus we recover the known result for the average subsystem purity Eq. \eqref{eq:QAE_averaged_purity}. Furthermore, by computing:
\begin{equation}
\label{eq:alpha_integral_goniometric}
 \braket{f^{L_A}}:=\int_{-\pi}^\pi\frac{d\alpha}{2\pi} f(\alpha)^{L_A} = \int_{-\pi}^\pi\frac{d\alpha}{2\pi} [\cos(\alpha/2)]^{2L_A} = \frac{(2L_A)!}{2^{2L_A}(L_A !)^2}\,,
\end{equation}
we obtain:
\begin{align}\label{eq:asym-global}
\mathbb{E}[\Delta S_A^{(2)}] &\simeq -\log\mathbb{E}[{\rm Tr}\rho_{A,Q}^2] + \log\mathbb{E}[{\rm Tr}\rho_{A}^2] \nonumber \\
&= \log \left[\frac{D_A + D_B -2 +I_2(\ket{\psi_0})(D-D_A-D_B+1)}{D_A \braket{f^{L_A}}+ D_B -2 +I_2(\ket{\psi_0})(D-D_A\braket{f^{L_A}}-D_B+1)}\right].
\end{align}

We can now study the asymptotic behaviour of $\mathbb{E}[\Delta S_A^{(2)}]$ as follows. Let us write:
\begin{equation}
\mathbb{E}[\Delta S_A^{(2)}] = \log  (1+r), \quad r=\frac{(1-\braket{f^{L_A}})(1-I_2(\ket{\psi_0})}{\braket{f^{L_A}} + 2^{L_B - L_A}-2^{1-L_A}+I_2(\ket{\psi_0})[2^{L_B}+2^{-L_A}-\braket{f^{L_A}}-2^{L_B-L_A}]}\,.
\end{equation}
We are interested in the regime in which $L,\,L_A\,L_B \gg 0$ and the ratios $L_A/L$, $L_B/L$ are fixed and between 0 and 1 as $L\to \infty$. Since $\braket{f^{L_A}} =\frac{1}{\sqrt{\pi L_A}}(1+O(L_A^{-1}))$ as $L_A \to \infty$, by substituting $L_B=L-L_A$ in the above expression, the ratio $r$ reads:
\begin{align}
r \simeq \frac{(1-\braket{f^{L_A}})(1-I_2(\ket{\psi_0})}{\braket{f^{L_A}} + 2^{L - 2L_A}+I_2(\ket{\psi_0}) 2^{L-L_A}}\,,
\end{align}
up to exponentially suppressed terms. Our goal is to find the subsystem size $L_A^*$ such that, in the limit of large $L$ and up to terms decaying exponentially in $L$, $\mathbb{E}[\Delta S_A^{(2)}]$ is zero when $L_A < L_A^*$, and it is an increasing function of $L_A$ when $L_A > L_A^*$. In other words, $L_A^*$ is the subsystem size at which symmetrization ceases to occur in the thermodynamic limit. We call $L_A^*$ the onset value of the asymmetry, or symmetrization scale. To find the latter, we recall that $0 < I_2(\ket{\psi_0})\le1$. Moreover, as we show below, we consider states for which $I_2(\ket{\psi_0})$ decays rapidly in $L$. Therefore, the onset value is obtained by imposing that $2^{L - 2L_A}+I_2(\ket{\psi_0}) 2^{L-L_A}$ is an exponentially decreasing function of $L_A$ for $L_A > L_A^*$. Recalling that the participation entropy is related to the participation ratio as $S_2^{\rm PE}(\ket{\psi_0})=-\log_2 I_2(\ket{\psi_0})$, this happens when:
\begin{equation}
\label{eq:asymmetry_onset_QAE}
\begin{cases}
&L - 2L_A < 0 \\
&L-L_A + \log_2 I_2(\ket{\psi_0}) <0
\end{cases}\quad \Rightarrow \quad L_A > L_A^* = \max\left\{\frac{L}{2}, L -S_2^{\rm PE}(\ket{\psi_0})\right\}\,.
\end{equation}
Equation \eqref{eq:asymmetry_onset_QAE} is our first main result: in the QAE, unlike the Haar random ensemble, the onset of the asymmetry can occur when the size of the subsystem $L_A$ is significantly larger than $L/2$, depending on the participation entropy of the initial state. In fact, with this definition, for $L_A> L_A^*$ we obtain at leading order in $L_A$: 
\begin{equation}
\label{eq:asymptotic_asymmetry_generic_QAE}
\mathbb{E}[\Delta S_A^{(2)}] \simeq \log\left[1+ \sqrt{\pi L_A}\left(1-\frac{1}{\sqrt{\pi L_A}}\right)(1-I_2(\ket{\psi_0}))\right]\,,
\end{equation}
whereas, for $L_A < L_A^*$, $\mathbb{E}[\Delta S_A^{(2)}] \simeq 0$, up to terms decaying exponentially in $L_A^*-L_A$. This means in particular that in the limiting case when the participation ratio reaches its maximum value $I_2(\ket{\psi_0})=1$, there will be no onset of the asymmetry.

 Let us now focus on two classes of initial states, for which $I_2(\ket{\psi_0})$, and thus $S_2^{\rm PE}(\ket{\psi_0})$, scales in different ways with the system size (see also the Supplemental material of \cite{szasz2026entanglement}):
\begin{enumerate}
    \item The homogeneous product states:
    \begin{equation}
\label{eq:def_homogeneous_product_state}
    \ket{\psi_0}=(\cos\theta \ket{0}+ \sin\theta\ket{1})^{\otimes L}\,, \quad \theta \in [0,\pi/2]\,,
    \end{equation}
    for which the second inverse participation ratio is:
    \begin{equation}
    \label{eq:I_2_product_state}
    I_2(\ket{\psi_0})=\left(1-\frac{1}{2}\sin^2(2\theta)\right)^L,
    \end{equation}
    i.e., for any $\theta \ne 0,\,\pi/2$, it is exponentially decreasing in the system size $L$,  thus $S_2^{\rm PE}(\ket{\psi_0})$ is linear in $L$. In particular, $2^{-L}\leq I_2(\ket{\psi_0})\leq 1$ with maximum value at $\theta=0,\,\pi/2$ (the initial state is a state of the computational basis, $\ket{\psi_0}=\ket{0}^{\otimes L}$ or $\ket{\psi_0}=\ket{1}^{\otimes L}$ and the state $\rho$ has trivially zero asymmetry for any realization in the QAE) and minimum value at $\theta=\frac{\pi}{4}$.
    \item The Dicke states:
    \begin{equation}\label{eq:dickestate-def}
    \ket{\psi_0}= \ket{D_k}:= \binom{L}{k}^{-1/2}\sum_{|s|=k}\ket{s}\,,
    \end{equation}
    where the sum is over the computational basis states $\ket{s}$ such that the bitstring $s$ has fixed Hamming weight $|s|=k$, i.e. it contains exactly $k$ 1's and $L-k$ 0's. The second inverse participation ratio is:
    \begin{equation}
    I_2(\ket{\psi_0}) = \binom{L}{k}^{-1},
    \end{equation}
    i.e. it has a power-law decay in the system size for $k\ne 0,\,L$: $I_2(\ket{D_k})=O(L^{-k})$.
\end{enumerate}
Of course, one could consider other states with different scaling behaviours of the participation entropy. For instance, the GHZ state $\ket{{\rm GHZ}} = (\ket{0}^{\otimes L} + \ket{1}^{\otimes L})/\sqrt{2}$ has a constant inverse participation ratio $I_2(\ket{{\rm GHZ}})=1/2$. However, the two examples considered here suffice to illustrate our main results.

\begin{figure}[h]
    \centering
    \includegraphics[width=0.47\linewidth]{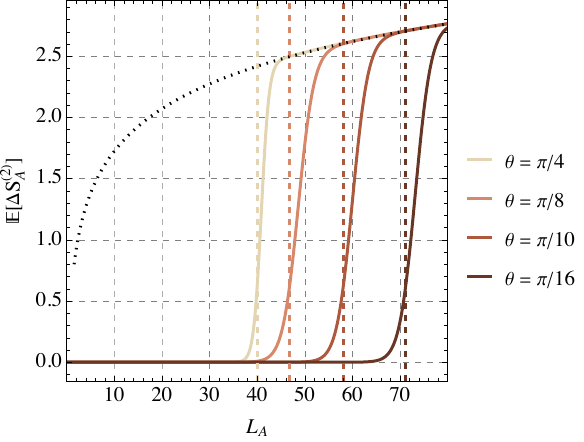}\quad\includegraphics[width=0.47\linewidth]{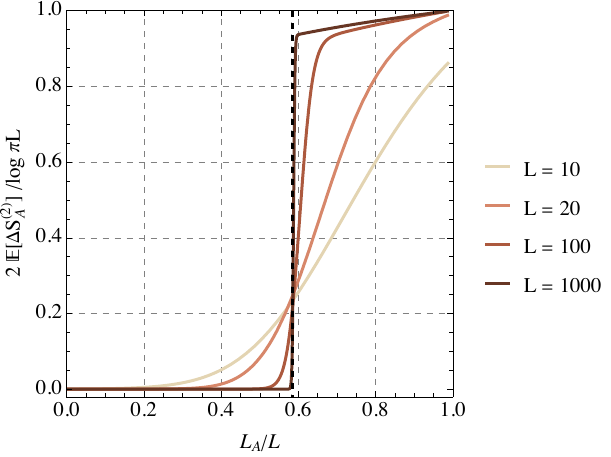}
    \caption{\textit{Left}: The entanglement asymmetry Page curve \eqref{eq:asym-global} averaged over the global quantum automaton ensemble starting from an initial product state  $\ket{\psi_0}$ \eqref{eq:def_homogeneous_product_state} for various values of the tilting angle $\theta$. The dashed lines denote the onset values $L_A^*$ as defined in Eq. \eqref{eq:asymmetry_onset_QAE}. The black dotted curve corresponds to $(1/2)\log(\pi L_A)$ as given in Eq.~\eqref{eq:asymptotic_asymmetry_typical}. We take $L = 80$. \textit{Right}: The finite size scaling of entanglement asymmetry Page curve \eqref{eq:asym-global} averaged over the global quantum automaton ensemble starting from an initial product state  $\ket{\psi_0}$ \eqref{eq:def_homogeneous_product_state} with $\theta = \pi/8$ for increasing system sizes $L$. The dashed line denote the onset value $L_A^* / L$ as defined in Eq. \eqref{eq:asymmetry_onset_QAE}.}
    \label{fig:asym-page-global}
\end{figure}

We now focus on the homogeneous product state. Up to terms decaying exponentially in $L$, Eq. \eqref{eq:asymptotic_asymmetry_generic_QAE} becomes
\begin{equation}
\label{eq:asymptotic_asymmetry_typical}
\mathbb{E}[\Delta S_A^{(2)}] \simeq  \frac{1}{2}\log(\pi L_A)\,,
\end{equation}
which reproduces the asymptotic behavior of the average R\'enyi-2 asymmetry in the Haar random ensemble when the subsystem size is $L_A > L/2$ \cite{BlackHole}. Moreover, from the expression Eq. \eqref{eq:I_2_product_state}, we can obtain the values of the tilting angle $\theta$ for which the onset of the asymmetry is at $L_A^* > L/2$. This happens when $\log_2 I_2(\ket{\psi_0})>-L/2$, that is when (for $0<\theta <\pi/2$):
\begin{equation}
0 < \theta < \frac{\arcsin{\sqrt{2-\sqrt{2}}}}{2}\,, \quad {\rm or} \quad \frac{\pi}{2} -\frac{\arcsin{\sqrt{2-\sqrt{2}}}}{2}< \theta < \frac{\pi}{2}\,.
\end{equation}
Note that $\frac{\arcsin{\sqrt{2-\sqrt{2}}}}{2} \simeq 0.436 \simeq \pi/7.21$. This phenomenology is quite surprising, as the specific scaling of the participation entropy for these states indicates that there is a critical angle below which the onset of the asymmetry changes from the typical value $L/2$ to $L-S_2^{\rm PE}(\ket{\psi_0})$. The averaged entanglement asymmetry as a function of the subsystem size for different homogeneous product states is shown in Fig. \ref{fig:asym-page-global}, together with the finite-size scaling of the results for a fixed value of the tilting parameter $\theta$. The values of $L_A^*$ as a function of $\theta$ are shown in Fig. \ref{fig:lastar}.

\begin{figure}
    \centering
    \includegraphics[width=0.47\linewidth]{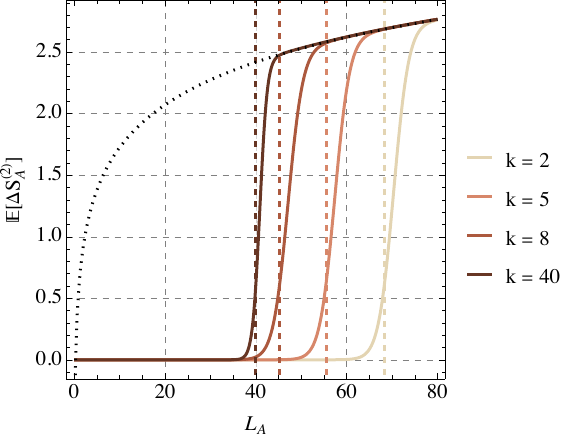}
    \caption{
    The entanglement asymmetry Page curve \eqref{eq:asym-global} averaged over the global quantum automaton ensemble starting from an initial Dicke-state  $\ket{D_k}$ \eqref{eq:dickestate-def} for different values of $k$. The dashed lines denote the onset values $L_A^*$ as defined in Eq. \eqref{eq:asymmetry_onset_QAE}. The black dotted curve corresponds to $(1/2)\log(\pi L_A)$ as given in Eq.~\eqref{eq:asymptotic_asymmetry_typical}. We take $L = 80$. }
    \label{fig:asym-page-global-dicke}
\end{figure}

\begin{figure}[h]
    \centering
    \includegraphics[width=0.37\linewidth]{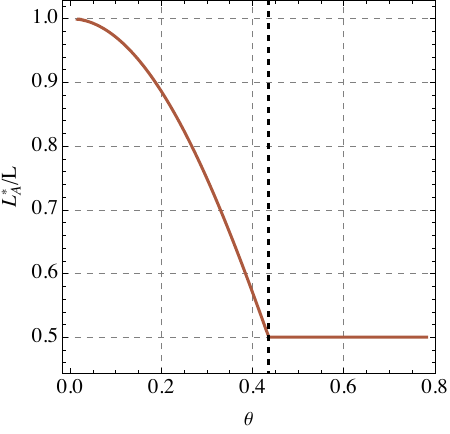} \includegraphics[width=0.376\linewidth]{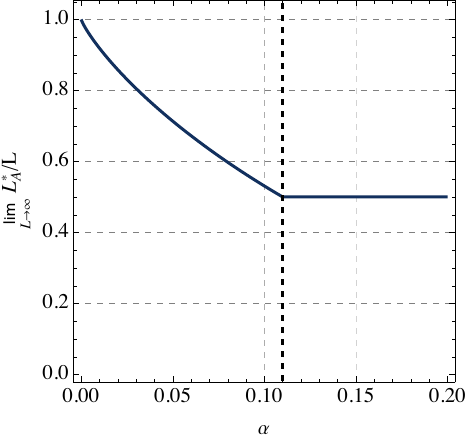}
    \caption{The critical value $L_A^*/L$ for different initial states. \textit{Left}: the critical value $L_A^*/L$ for initial product states \eqref{eq:def_homogeneous_product_state} computed from Eq. \eqref{eq:asymmetry_onset_QAE} using $L = 80$. The black dashed line denotes to $\theta^* = \frac{\arcsin{\sqrt{2-\sqrt{2}}}}{2}$, corresponding to the smallest value where $L_A^*/L = 1/2$. \textit{Right}: the critical value $\lim_{L\rightarrow\infty}L_A^*/L$ for initial Dicke states, computed from Eq. \eqref{eq:dicke-criticalL}, corresponding to infinite system size. The black dashed line denotes $\alpha^* \approx 0.11$.}
    \label{fig:lastar}
\end{figure}

For the Dicke state $\ket{D_k}$, Eq. \eqref{eq:asymptotic_asymmetry_generic_QAE} yields the same leading order term of Eq. \eqref{eq:asymptotic_asymmetry_typical}, but with a power-law decaying correction $O(L^{-k})$. The averaged entanglement asymmetry Page-curve for different values of $k$ and $L = 80$ is shown in Fig. \ref{fig:asym-page-global-dicke}. Interestingly, unlike the case of homogeneous product states where the critical angle $\theta$ does not depend on the system size, for the states $\ket{D_k}$ the critical value of $k$ at which the onset value shifts from the typical value $L/2$ to $L-S_2^{\rm PE}(\ket{\psi_0})$ depends on the system size. This is due to the different scaling of the participation entropy with the system size $L$. In the large $L$ limit, letting $k =\alpha L$, we can approximate
\begin{equation}\label{eq:dicke-criticalL}
\lim_{L\rightarrow \infty} \frac{1}{L}(L+\log_2 I_2(\ket{\psi_0})) = \lim_{L\rightarrow \infty} \left(1 - \frac{\log_2\binom{L}{\alpha L}}{L}\right) = \lim_{L\rightarrow \infty} \left[1 - H_2(\alpha) + \frac{\log_2 L }{2L} + O\left(\frac{1}{L}\right)\right] = 1 - H_2(\alpha)\,,
\end{equation}
where $H_2(\alpha) = -\alpha \log_2 \alpha - (1-\alpha) \log_2(1-\alpha)$ is the binary entropy and we used Stirling's formula $\binom{L}{\alpha L} \sim \frac{2^{L H_2(\alpha)}}{\sqrt{2\pi L \alpha(1-\alpha)}}$. Restricting ourselves to values of $\alpha \leq 1/2$, we arrive at the critical value of $k$ for which $L-S_2^{\rm PE}(\ket{\psi_0})> L/2$:
\begin{equation}
    k < \alpha^* L \approx 0.11 L\,.
\end{equation}
That is, the shift in the symmetrization scale happens for Dicke states $\ket{D_k}$ with $k < \alpha^* L$, as shown in Fig. \ref{fig:lastar}.

\section{Dynamics of entanglement asymmetry in a two-local quantum automaton circuit}
\label{sec:Dynamics of entanglement asymmetry in a two-local quantum automaton circuit}

In this Section, we study the dynamics of the averaged R\'enyi-2 entanglement asymmetry under a non-local quantum circuit evolution, starting from a homogeneous product state. In particular, we consider a 2-local quantum circuit, that is, at any time step, with a certain probability a 2-body gate $U_{j,k}$, consisting of a permutation of basis states followed by a global phase transformation is applied to two qubits in random sites $j$ and $k$. A slight generalization of this protocol was employed in \cite{piroli2020random} to study the dynamics of the averaged R\'enyi-2 entropy under a 2-local random unitary circuit (RUC). The same protocol we study here was followed in \cite{Ares25} to compute the averaged R\'enyi-2 asymmetry in the Haar random case, and the averaged R\'enyi-2 entropy in the case of random permutation circuits and random quantum automata in \cite{szasz2026entanglement}. 

The advantage with respect to a local (brickwork) circuit, which requires mapping the system to a classical statistical model to obtain analytical predictions \cite{nahum2017quantum,nahum2018operator,fisher2023random,Ares25}, is that with a non-local protocol one can construct a set of observables with a Markovian evolution, governed by a system of differential equations. By numerically solving these differential equations, the expectation values of the charged moments at any time $t>0$ for all subsystem sizes are obtained at once, as well as the expectation values of the subsystem purities. As we will discuss in Section \ref{sec:Connection to subsystem coherence}, from the same system of differential equations one can also extract the average coherence of all subsystems. 

\subsection{Differential equations for the charged moments}
In a 2-local quantum circuit, at each discrete time step $t$, with a certain probability $p$ we apply a two-body gate $U_{j,k}$, $1\le j  < k \le L$, to the state $\rho(t)$, so that at time $t+\Delta t$ the state is $\rho(t+\Delta t)= U(t+\Delta t)\rho(t)U^\dagger(t + \Delta t)$, where $U(t+\Delta t) = U_{j,k} U(t)$ with probability $p=L \Delta t$, and $U(t+\Delta t) = U(t)$ with probability $1-p$. At time $t=0$, $U(0)$ is the identity operator. Then, we average over the sites $j,k$ (chosen with uniform probability) and over the ensemble from which the operators $U_{j,k}$ are drawn, that is the quantum automaton circuit ensemble (QAC). In the continuous time limit $\Delta t \to 0$ (see Appendix A of \cite{piroli2020random}), this evolution protocol yields a master equation for the two-replica average $\mathbb{E}_{\rm QAC}\ket{\rho(t)\otimes\rho(t)}$:
\begin{equation}
\label{eq:Lindblad_equation}
  \frac{d}{dt}\mathbb{E}\ket{\rho(t)\otimes\rho(t)}= - \mathcal{L}\,\mathbb{E}\ket{\rho(t)\otimes\rho(t)}\,,
\end{equation}
with
\begin{equation}
    \label{eq:Lindbladian}
    \mathcal{L} = \frac{2}{L-1}\sum_{1\leq j< k\leq L}(1-\mathcal{U}_{j,k}),\quad \mathcal{U}_{j,k} = \mathbb{E}\left[U^{*}_{j,k}\otimes U_{j,k} \otimes U^{*}_{j,k} \otimes U_{j,k}\right]\,.
\end{equation}
The expectation value $\mathcal{U}_{j,k}$ can be obtained in the very same way as the corresponding expectation value in the global ensemble QAE (see Appendix \ref{sec:Choi representation for random quantum automata}), with the only difference that now $\mathcal{U}_{j,k}$ is an operator acting non-trivially only on the replica Hilbert space $(\mathcal{H}_j\otimes \mathcal{H}_k)^{\otimes 4}$. Thus, by replacing $D=2^L$ in Eq. \eqref{eq:second_moment_QAE_main} with $D=4$, we have:
\begin{align}
&\mathbb{E}_{\rm QAC}\left[U^{*}_{j,k}\otimes U_{j,k} \otimes U^{*}_{j,k} \otimes U_{j,k}\right] = \nonumber \\ &\frac{\ket{I^+}_{j,k}\bra{I^+}_{j,k} + \ket{I^-}_{j,k}\bra{I^-}_{j,k} + 5\ket{I^0}_{j,k}\bra{I^0}_{j,k}-(\ket{I^-}_{j,k}\bra{I^0}_{j,k}+\ket{I^0}_{j,k}\bra{I^-}_{j,k}+\ket{I^+}_{j,k}\bra{I^0}_{j,k}+\ket{I^0}_{j,k}\bra{I^+}_{j,k})}{12}\,,
\end{align}
where we introduced the notation $\ket{I^a}_{j,k}=\ket{I^a}_j\ket{I^a}_k$ for $a=-,+,0$.

The averaged time-dependent charged moment $Z_A^{(2)}(\alpha,t)$ is given by:
\begin{equation}
\label{eq:charged_moment_QAC}
\mathbb{E}[Z_A^{(2)}(\alpha,t)]={}_A\bra{\mathcal{I}^-_\alpha}{}_B\bra{\mathcal{I}^+}\mathbb{E}\ket{\rho(t)\otimes\rho(t)}\,,
\end{equation}
and its time derivative is:
\begin{equation}
\label{eq:time_derivative_Z_2}
\frac{d\mathbb{E}[Z_A^{(2)}(\alpha,t)]}{dt}={}_A\bra{\mathcal{I}^-_\alpha}{}_B\bra{\mathcal{I}^+}(-\mathcal{L})\mathbb{E}\ket{\rho(t)\otimes\rho(t)}\,.
\end{equation}
If the right-hand side of the above equation was a linear combination of the averaged moments $Z_{A'}^{(2)}(\alpha,t)$ for different subsystems $A'$, one would have a complete system of equations for the time evolution of the charged moments of all subsystems. However, this is not the case, because the action of $\mathcal{U}_{j,k}$ on ${}_A\bra{\mathcal{I}^-_\alpha}{}_B\bra{\mathcal{I}^+}$ produces extra terms other than a linear combination of states ${}_{A'}\bra{\mathcal{I}^-_\alpha}{}_{B'}\bra{\mathcal{I}^+}$ for different (complementary) subsystems $A'$ and $B'$, even if one considers a simpler RUC \cite{Ares25}. In order to obtain a closed set of equations, we notice that the Lindbladian $\mathcal{L}$ is invariant under site permutation. If one also chooses a permutation-invariant initial state, such as the homogeneous product states Eq. \eqref{eq:def_homogeneous_product_state}, then the (averaged) state at any $t>0$ is site-permutation invariant. In this scenario, a closed set of equations is obtained by introducing the following states:
\begin{equation} 
\label{eq:G_state_definition}
\ket{G_{n_\alpha, n_-, n_+, n_0}^\alpha}:= \frac{1}{L!} \sum_{\pi \in {\mathcal{S}}_L}  \hat{\pi} (\ket{I_\alpha^-}^{\otimes n_\alpha} \otimes \ket{I^-}^{\otimes n_-} \otimes\ket{I^+}^{\otimes n_+} \otimes \ket{I^0}^{\otimes n_0})\,.
\end{equation} 
In the above expression, the indices $n_\alpha$, $n_-$, $n_+$ and $n_0$ are non-negative integers such that $n_\alpha + n_- + n_+ + n_0=L$, $\pi$ is a permutation of the $L$ sites and $\hat{\pi}$ implements that permutation on the states. Thus, $\ket{G_{n_\alpha, n_-, n_+, n_0}^\alpha}$ is the permutation-invariant state in the two-replica space containing $n_\alpha$ copies of the state $\ket{I^-_\alpha}_j$ and so on for the other indices. This is a generalization of the states introduced in \cite{szasz2026entanglement} to compute the averaged purity in non-local QACs, and of the states introduced in \cite{Ares25} to compute the averaged asymmetry in non-local RUCs.

It is easy to derive the action of the operators $\mathcal{U}_{j,k}$ on any state. The following relations can be found in \cite{szasz2026entanglement}:
\begin{align}
\label{eq:QAC_overlaps_old}
\mathcal{U}_{j,k}\ket{I^\pm}_j \ket{I^\pm}_k &=\ket{I^\pm}_j\ket{ I^\pm}_k\,, \nonumber \\
\mathcal{U}_{j,k}\ket{I^0}_j\ket{ I^0}_k &=\ket{I^0}_j\ket{ I^0}_k\,,\nonumber \\
\mathcal{U}_{j,k}\ket{I^\pm }_j\ket{I^0}_k &=\frac{1}{3}\left(\ket{I^\pm}_j\ket{ I^\pm}_k+ 2 \ket{I^0}_j\ket{ I^0}_k \right)\,,\nonumber \\
\mathcal{U}_{j,k}\ket{I^+}_j\ket{ I^-}_k &=\frac{1}{3}\left(\ket{I^-}_j \ket{I^-}_k + \ket{I^+}_j\ket{ I^+}_k + \ket{I^0}_j\ket{ I^0}_k \right)\,.
\end{align} 
We then need to evaluate the action of $\mathcal{U}_{j,k}$ on the states with at least one factor $\ket{I^-_\alpha}_j$. This is easily done by using the overlaps in Eqs. \eqref{eq:I_alpha_overlaps}, and we obtain:
\begin{align}
\label{eq:QAC_overlaps_alpha}
\mathcal{U}_{j,k}\ket{I^-_\alpha}_j \ket{I^-_\alpha}_k &=A(\alpha)\ket{I^-}_j\ket{ I^-}_k + B(\alpha)\ket{I^0}_j\ket{ I^0}_k\,, \nonumber \\
\mathcal{U}_{j,k}\ket{I^-_\alpha}_j \ket{I^+}_k &=\frac{1}{3}(\ket{I^+}_j\ket{ I^+}_k+C(\alpha)\ket{I^-}_j\ket{ I^-}_k+D(\alpha)\ket{I^0}_j\ket{ I^0}_k)\,,\nonumber \\
\mathcal{U}_{j,k}\ket{I^-_\alpha}_j \ket{I^0}_k &=\frac{1}{3}(C(\alpha)\ket{I^-}_j\ket{ I^-}_k+ E(\alpha) \ket{I^0}_j\ket{ I^0}_k )\,,\nonumber \\
\mathcal{U}_{j,k}\ket{I^-_\alpha}_j \ket{I^-}_k &=\frac{1}{3}(F(\alpha)\ket{I^-}_j \ket{I^-}_k + G(\alpha)\ket{I^0}_j\ket{ I^0}_k) \,,
\end{align} 
with 
\begin{align}
A(\alpha) &=\frac{4 f(\alpha)^2-1}{3}\,, \quad B(\alpha) =\frac{4}{3}(1-f(\alpha)^2)\,, \quad C(\alpha)=2f(\alpha)-1\,, \quad D(\alpha)=3-2f(\alpha)\,, \nonumber \\
E(\alpha) &=2(2-f(\alpha))\,, \quad F(\alpha) =4 f(\alpha)-1\,, \quad G(\alpha)=4(1-f(\alpha)^2)\,.
\end{align}
Since $f(\alpha)=\cos^2(\alpha/2)$, $f(0)=1$ and Eqs. \eqref{eq:QAC_overlaps_alpha} reduce to \eqref{eq:QAC_overlaps_old} when $\alpha=0$. 

Following \cite{szasz2026entanglement}, we want to write a system of ODEs for the functions
\begin{equation} 
G_{n_\alpha, n_-, n_+, n_0}^\alpha(t):=\langle G_{n_\alpha, n_-, n_+, n_0}^\alpha| \mathbb{E}\ket{\rho(t)\otimes \rho(t)}\,,
\end{equation} 
with the convention that $G_{n_\alpha, n_-, n_+, n_0}^\alpha$ is zero whenever any of the indices $n_\alpha,\, n_-,\, n_+,\, n_0$ is smaller than zero or larger than $L$. Since there are $\binom{L+3}{L}$ of these functions, corresponding to the number of tuples of non-negative integers $(n_\alpha,n_-,n_+,n_0)$ such that $n_\alpha + n_- + n_+ + n_0=L$, the number of independent equations in the system is $\binom{L+3}{L} \sim L^3$. These are obtained from the Lindbladian operator in Eq. \eqref{eq:Lindbladian}:
\begin{equation} 
\frac{d G_{n_\alpha, n_-, n_+, n_0}^\alpha(t)}{dt}= \langle G_{n_\alpha, n_-, n_+, n_0}^\alpha|(-\mathcal{L}) \, \mathbb{E}\ket{\rho(t)\otimes \rho(t)}\,,
\end{equation} 
together with Eqs. \eqref{eq:QAC_overlaps_old} and \eqref{eq:QAC_overlaps_alpha}. Leaving the details of the derivation to Appendix \ref{sec:Details on the dynamics in non-local QAC}, we obtain:
\begin{equation}
\label{eq:G_system_short}
\frac{d G_{n_\alpha, n_-, n_+, n_0}^\alpha(t)}{dt} = \sum_{j,k,r=-2}^2 M^\alpha_{j,k,r} \, G_{n_\alpha+j,\, n_- + k,\, n_+ + r,\, n_0-(j+k+r)}^\alpha(t)\,,
\end{equation}
with
\begin{align} 
M^\alpha_{0,0,0}&=\frac{n_\alpha + n_-^2 + n_+^2 + n_0^2 -L^2}{L-1}\,, \quad
M^\alpha_{-2,2,0}=\frac{n_\alpha(n_\alpha-1)}{L-1}A(\alpha)\,, \quad M^\alpha_{-2,0,0}=\frac{n_\alpha(n_\alpha-1)}{L-1}B(\alpha)\,, \nonumber \\
M^\alpha_{-1,1,0}&=\frac{2n_\alpha n_-}{3(L-1)}F(\alpha)\,, \quad M^\alpha_{-1,-1,0}=\frac{2n_\alpha n_-}{3(L-1)}G(\alpha)\,, \nonumber \\
M^\alpha_{-1,0,1}&=\frac{2n_\alpha n_+}{3(L-1)}\,, \quad M^\alpha_{-1,2,-1}=\frac{2n_\alpha n_+}{3(L-1)}C(\alpha)\,, \quad M^\alpha_{-1,0,-1}=\frac{2n_\alpha n_+}{3(L-1)}D(\alpha)\,,\nonumber\\
M^\alpha_{-1,2,0}&=\frac{2n_\alpha n_0}{3(L-1)}C(\alpha)\,, \quad M^\alpha_{-1,0,0}=\frac{2n_\alpha n_0}{3(L-1)}E(\alpha)\,,\quad 
M^\alpha_{0,-1,1} =M^\alpha_{0,1,-1}=M^\alpha_{0,-1,-1}=\frac{2n_- n_+}{3(L-1)}\,,\nonumber\\
M^\alpha_{0,1,0}&=\frac{2n_- n_0}{3(L-1)}\,, \quad M^\alpha_{0,-1,0}=\frac{4n_- n_0}{3(L-1)}\,, \quad
M^\alpha_{0,0,1}=\frac{2n_+ n_0}{3(L-1)}\,, \quad M^\alpha_{0,0,-1}=\frac{4n_+ n_0}{3(L-1)}\,,
\end{align} 
the only non-vanishing coefficients. Thus, all the functions $G_{n_\alpha, n_-, n_+, n_0}^\alpha(t)$ evolve with a closed system of linear ODEs. In particular, for permutation-invariant initial states, the average charged moments of a subsystem $A$ are functions only of the subsystem size $L_A$. Indeed, we can rewrite Eq. \eqref{eq:charged_moment_QAC} as:
\begin{equation}
\mathbb{E}[Z_A^{(2)}(\alpha,t)]= \frac{1}{L!}\sum_{\pi \in \mathcal{S}_L}\hat{\pi}(\bra{I^-_\alpha}^{\otimes L_A}\otimes \bra{I^-_\alpha}^{\otimes (L-L_A)})\mathbb{E}\ket{\rho(t)\otimes \rho(t)}=G^\alpha_{L_A,0,L-L_A,0}(t)=:\mathbb{E}[Z_{L_A}^{(2)}(\alpha,t)]\,.
\end{equation}
The evolution equation for $\mathbb{E}[Z_{L_A}^{(2)}(\alpha,t)]$ can be read-off from Eq. \eqref{eq:G_system_short}:
\begin{align}
\frac{d \mathbb{E}[Z_{L_A}^{(2)}(\alpha,t)]}{dt}&= \frac{L_A(1+L_A-2L)}{L-1}\mathbb{E}[Z_{L_A}^{(2)}(\alpha,t)] \nonumber \\ &+ \frac{L_A(L_A-1)}{L-1}[A(\alpha) G^\alpha_{L_A-2,2,L-L_A,0}(t) + B(\alpha) G^\alpha_{L_A-2,0,L-L_A,2}(t)] \nonumber \\
&+\frac{2L_A(L-L-L_A)}{3(L-1)}[\mathbb{E}[Z_{L_A-1}^{(2)}(\alpha,t)]+C(\alpha) G^\alpha_{L_A-1,2,L-L_A-1,0}(t) + D(\alpha) G^\alpha_{L_A-1,0,L-L_A-1,2}(t)]\,.
\end{align}
We notice that, when $n_\alpha=0$ the equations \eqref{eq:G_system_short} reduce to those derived in \cite{szasz2026entanglement} for the evolution of the average subsystem purity in a QAC, i.e. a complete system of equations for the functions $G_{n_-,n_+,n_0}(t) := G^{\alpha=0}_{n_\alpha=0,n_-,n_+,n_0}(t)$. Remarkably, by solving this system, one obtains as a byproduct the dynamics of the relative entropy of coherence for all subsystems. We show how this is done in Appendix \ref{sec:Details on the dynamics in non-local QAC}. 

We now need to establish the initial conditions for the functions $G_{n_\alpha,n_-, n_+, n_0}^\alpha(t)$, i.e.:
\begin{equation}
G^\alpha_{n_\alpha,n_-,n_+,n_0}(0) = \braket{G^\alpha_{n_\alpha,n_-,n_+,n_0}|\rho_0\otimes \rho_0}\,,
\end{equation}
where $\rho_0=\ket{\psi_0}\bra{\psi_0}$ and $\ket{\psi_0}$ is the homogeneous product state Eq. \eqref{eq:def_homogeneous_product_state}. It follows from the definition Eq. \eqref{eq:CJ_mapping_main} that we can write (see also the Supplemental material of \cite{szasz2026entanglement}):
\begin{equation} 
\ket{\rho_0\otimes \rho_0} =\bigotimes_{j=1}^L \sum_{a,b,c,d=0}^{1} \lambda_a^*\lambda_b \lambda_c^* \lambda_d \ket{abcd}_j\,, \quad \lambda_0:= \cos \theta,\, \lambda_1=\sin \theta\,.
\end{equation} 
Since $\rho_0$ is a permutation invariant state, one easily obtains:
\begin{align} 
\label{eq:hom_product_state_initial_conditions_G_functions}
G^\alpha_{n_\alpha,n_-,n_+,n_0}(0)
&= \left(\bra{I^-_\alpha} \sum_{a,b,c,d=0}^{1} \lambda_a^*\lambda_b \lambda_c^* \lambda_d \ket{abcd}\right)^{n_\alpha} 
\left(\bra{I^-} \sum_{a,b,c,d=0}^{1} \lambda_a^*\lambda_b \lambda_c^* \lambda_d \ket{abcd} \right)^{n_-} 
\nonumber\\
& \times \left(\bra{I^+} \sum_{a,b,c,d=0}^{1} \lambda_a^*\lambda_b \lambda_c^* \lambda_d \ket{abcd} \right)^{n_+} \left(\bra{I^0} \sum_{a,b,c,d=0}^{1} \lambda_a^*\lambda_b \lambda_c^* \lambda_d \ket{abcd} \right)^{n_0}\nonumber\\
&= \left(1-\sin^2\frac{\alpha}{2} \sin^2(2\theta)  \right)^{n_\alpha} \left(1-\frac{1}{2}\sin^2(2\theta) \right)^{n_0}\,.
\end{align} 
As a sanity check, let us consider some specific initial conditions in detail. When $n_\alpha=n_0=0$, one has ${\rm Tr}\rho_{0,A}^2=G^{0}_{0,L_A,L-L_A,0}(0)=1$ as expected, since the initial state is a product state. On the other hand, the quantity $G^{0}_{0,0,0,L}(0)$ reproduces the second inverse participation ratio of $\ket{\psi_0}$, Eq. \eqref{eq:I_2_product_state}. Finally, Although the time-dependent averaged asymmetry can only be obtained numerically by integrating $\mathbb{E}[Z_A^{(2)}(\alpha,t)]$ over $\alpha$, the R\'enyi-2 asymmetry of $\ket{\psi_0}$ can be obtained in the following way:
\begin{align}
\label{eq:explicit_asymmetry_initial_hom_product_state}
\Delta S^{(2)}_A(t=0)&=-\log\left[\int_{-\pi}^{\pi} \frac{d\alpha}{2\pi} Z_{L_A}^{(2)}(\alpha,0)\right]=-\log\left[\int_{-\pi}^{\pi} \frac{d\alpha} {2\pi}\left(1-\sin^2\frac{\alpha}{2}\sin^2(2\theta)\right)^{L_A}\right] \nonumber \\
&=-\log\left[\sum_{k=0}^{L_A}\binom{L_A}{k}(-1)^k\sin^{2k}(2\theta)\frac{(2k)!}{2^{2k}(k!)^2}\right]=-\log\left[{}_2 F_1\left(-L_A,\frac{1}{2},1; \sin^2(2\theta)\right)\right]\,,
\end{align}
where we used the result Eq. \eqref{eq:alpha_integral_goniometric}, which is the same when the function $\cos(\alpha/2)$ is replaced by $\sin(\alpha/2)$ in the integrand, and the standard definition of the Hypergeometric function \cite{abramowitz1966handbook}. The above result \eqref{eq:explicit_asymmetry_initial_hom_product_state} was first obtained, for a generic number $n$ of replicas, in \cite{ares2023entanglement}.

We conclude by observing that, even though finding the analytical solution to the system of equations \eqref{eq:G_system_short} is extremely challenging, the stationary solutions to the latter, obtained by imposing $d G_{n_\alpha, n_-, n_+, n_0}^\alpha(t)/dt=0$, are much easier to find. Indeed, these are given by the following functions:
\begin{equation}
\mathcal{G}^\alpha_{n_\alpha, n_-,n_+,n_0}:= \langle G_{n_\alpha, n_-, n_+, n_0}^\alpha|\mathbb{E}_{\rm QAE}[U^*\otimes U\otimes U^*\otimes U]\ket{\rho_0 \otimes \rho_0}\,.
\end{equation}
The expectation value in the above expression is taken in the global QAE, and indeed the quantities $\mathcal{G}^\alpha_{n_\alpha, n_-,n_+,n_0}$ are simple generalizations of the expression Eq. \eqref{eq:averaged_charged_moment_QAE}, and they can be computed with the same techniques used in Section \ref{sec:Averaged entanglement asymmetry in the quantum automaton ensemble}. Explicitly: 
\begin{equation}
 \mathcal{G}^\alpha_{n_\alpha, n_-,n_+,n_0} = \frac{D_\alpha f(\alpha)^{n_\alpha}+D_+-2 + I_2(\ket{\psi_0})[D+1-D_\alpha f(\alpha)^{n_\alpha}-D_+]}{D-1}\,, 
\end{equation}
where $D_{\alpha/-/+/0}=2^{n_\alpha/n_-/n_+/n_0}$ and $D=D_\alpha D_-D_+D_0$.
A direct substitution into Eq. \eqref{eq:G_system_short} shows that these functions yield a steady-state solution of the system. This proves that the ensemble averages of the R\'enyi-2 entanglement asymmetry computed in $\mathcal{E}_{\rm QAC}(\infty)$ and $\mathcal{E}_{\rm QAE}$ coincide, at any finite size $L$. 
\subsection{Numerical results}

\begin{figure}[h]
    \centering
    \includegraphics[width=1.\linewidth]{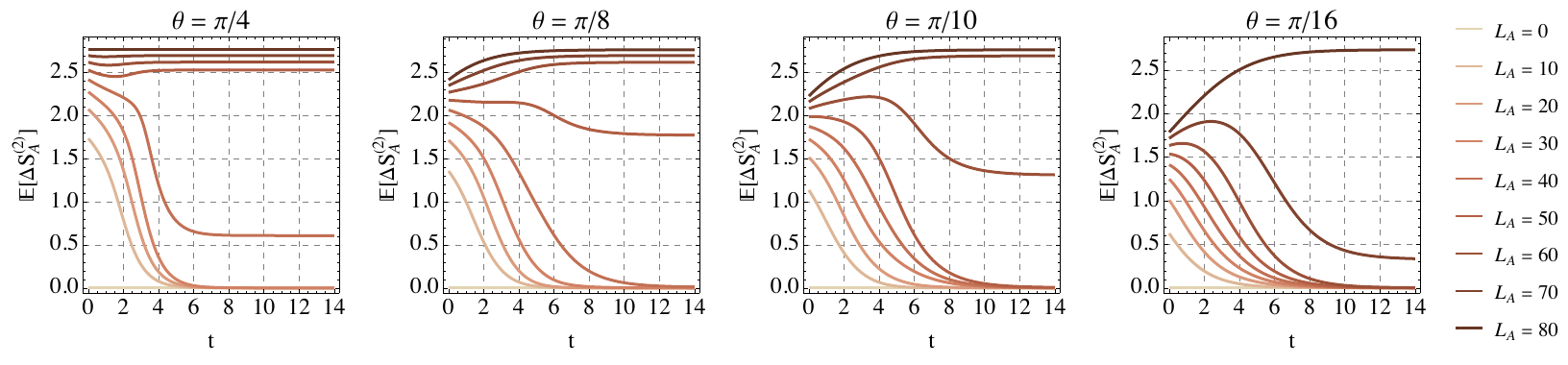}
    \caption{The averaged entanglement asymmetry under 2-local random quantum automaton circuit evolution starting from an initial product state $\ket{\psi_0}$ \eqref{eq:def_homogeneous_product_state} for various values of the tilting angle $\theta$. We take $L = 80$.}
    \label{fig:asym-dynamics}
\end{figure}

\begin{figure}[h]
    \centering
    \includegraphics[width=0.47\linewidth]{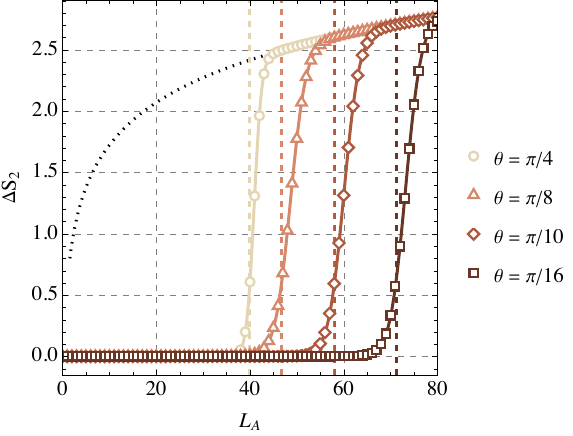}
    \caption{The entanglement asymmetry Page curve averaged over the 2-local quantum automaton ensemble (markers) starting from an initial product state  $\ket{\psi_0}$ \eqref{eq:def_homogeneous_product_state} for various values of the tilting angle $\theta$. The global ensemble results are shown in solid lines. The dashed lines denote the onset values $L_A^*$ as defined in Eq. \eqref{eq:asymmetry_onset_QAE}. The black dotted curve corresponds to $(1/2)\log(\pi L_A)$. We take $L = 80$.}
    \label{fig:asym-page-local}
\end{figure}

We now discuss the time evolution of the averaged entanglement asymmetry under the 2-local random quantum automaton circuit evolution, shown in Fig. \ref{fig:asym-dynamics}. The results are obtained through numerical integration of Eq. \eqref{eq:G_system_short}, following the outline of this section. The system is initialized in a homogenous product state \eqref{eq:def_homogeneous_product_state} for different values of the tilting angle $\theta$. We set the system size $L = 80$, and run the evolution until the asymmetry reaches their stationary value. As apparent from Fig. \ref{fig:asym-dynamics}, curves corresponding to different subsystem sizes $L_A$ relax to either a finite value or zero, depending on the specific choice of $L_A$ and $\theta$. The extracted stationary values are plotted against the subsystem size in Fig. \ref{fig:asym-page-local}, together with the global results \eqref{eq:asym-global}. As expected, the long-time local results match the global ones: the asymmetry stays zero for subsystem sizes $L_A\leq L_A^*$, then rapidly increase to the value $(1/2)\log(\pi L_A)$. The onset point $L_A^*$ is the same as for the global case, given in Eq. \eqref{eq:asymmetry_onset_QAE}.

\section{Connection to subsystem coherence}
\label{sec:Connection to subsystem coherence} 
In this section, we provide a justification for the results presented in the previous sections. In particular, we give a physical argument as to why in the QAE the onset of the averaged asymmetry may occur at subsystem sizes much larger than $L/2$.  Essentially, we show that a very small participation entropy (equivalently, a very high degree of localization in the Hilbert space) prevents the build-up of subsystem coherence up to a subsystem size precisely given by Eq. \eqref{eq:asymmetry_onset_QAE}. This complements the picture presented in \cite{szasz2026entanglement}, where it was shown that a small participation entropy  prevents entanglement saturation to a volume law at large subsystem sizes. First, we discuss the distinguishability between the subsystem's density matrix and the infinite-temperature state in the ensemble of random quantum automata. Namely, we generalize the decoupling inequality first derived in \cite{hayden2007black}. The decoupling inequality explains why symmetrization occurs at small subsystem sizes, but it is not enough to infer the value $L_A^*$ in Eq. \eqref{eq:asymmetry_onset_QAE}. Therefore, we propose a criterion for the onset of asymmetry based on the notion of subsystem coherence and use it to explain the findings of the previous sections.

\subsection{The decoupling inequality}
We prove an upper bound (decoupling inequality) on the averaged squared trace distance between the state $\rho_A$ and the maximally mixed state on $\mathcal{H}_A$, where the average is over $\mathcal{E}_{\rm QAE}$. Let us recall some preliminary definitions (see e.g. \cite{watrous2018theory}). Let $\mathcal{L}(\mathcal{H})$ be the space of linear operators over $\mathcal{H}$, and $p\ge1$. The Schatten $p$-norm of $A \in \mathcal{L}(\mathcal{H})$ is defined as:
\begin{equation}
\label{eq:def_Schatten_p_norm}
||A||_p := [{\rm Tr}(A^\dagger A)^{p/2}]^\frac{1}{p}\,.
\end{equation}
We are interested in Schatten norms with $p=1,\,2$. For $p=1$ this is the trace norm, which coincides with the sum of the singular values of $A$: $||A||_1= {\rm Tr}\sqrt{A^\dagger A}=\sum_k |\lambda_k(A)|$. For $p=2$, the Schatten norm is the Hilbert-Schmidt (HS) norm, induced by the HS inner product $\langle A, B\rangle_{\rm HS}={\rm Tr}(B^\dagger A)$. The Schatten norm satisfies the following H\"older inequality. For every $A,B \in \mathcal{L}(\mathcal{H})$:
\begin{equation}
\label{eq:Holder_inequality}
||A B||_1 \le ||A||_p ||B||_q, \quad {\rm for} \quad \frac{1}{p}+\frac{1}{q}=1.
\end{equation}
In particular, take $p=q=2$ in the above inequality, and take $B=\mathds{1}_{\mathcal{H}}$. Then, since ${\rm dim}\mathcal{H}=D$, $||\mathds{1}||_2=\sqrt{{\rm Tr}_{\mathcal{H}}\,\mathds{1}}=\sqrt{D}$. Thus, the Holder inequality reads:
\begin{equation}
\label{eq:Holder_p=2}
||A||_1 \le \sqrt{D} ||A||_2\,.
\end{equation}

The decoupling inequality is easily obtained from the inequality \eqref{eq:Holder_p=2}. In fact, considering a bipartite Hilbert space as in \eqref{def:bipartite_Hilbert_space}, in order to bound the trace distance between $\rho_A$ and the completely mixed state  $\mathds{1}_{A}/D_A$, we define $X:= \rho_A -\mathds{1}_{A}/D_A$ and apply the above inequality (squared) to $X$:
\begin{equation}
\left |\left | \rho_A - \frac{\mathds{1}_{A}}{D_A}  \right | \right |_1^2 \le D_A \left |\left | \rho_A - \frac{\mathds{1}_{A}}{D_A}  \right | \right |_2^2 = D_A \left[{\rm Tr} \rho_A^2-\frac{1}{D_A}\right]\,.
\end{equation}
The above decoupling inequality holds in full generality. Suppose that $\rho = U\ket{\psi_0} \bra{\psi_0}U^\dagger$, for some initial state $\ket{\psi_0}$, and the unitary matrix $U\in U(D)$ is drawn from some ensemble $\mathcal{E}$. Then:
\begin{equation}
\mathbb{E}\left |\left | \rho_A - \frac{\mathds{1}_{A}}{D_A}  \right | \right |_1^2 \le D_A \left[\mathbb{E}({\rm Tr} \rho_A^2)-\frac{1}{D_A}\right]\,.
\end{equation}
For the Haar random ensemble, from Eq. \eqref{eq:Haar_random_average_purity} it follows
\begin{equation}
\mathbb{E}_{\rm Haar}\left |\left | \rho_A - \frac{\mathds{1}_A}{D_A}  \right | \right |_1^2 \le \frac{D_A^2 -1}{D_A D_B + 1} = \frac{2^{2L_A}-1}{2^L +1} < 2^{L_A-L_B}\,,
\end{equation}
which is the inequality presented in \cite{BlackHole,Ares25}. 

On the other hand, for $\mathcal{E}_{\rm QAE}$, from Eq. \eqref{eq:QAE_averaged_purity} it follows that:
\begin{equation}
\mathbb{E}_{\rm QAE}\left |\left | \rho_A - \frac{\mathds{1}_{A}}{D_A}  \right | \right |_2^2 = \frac{D_A + D_B -2 + I_2(\ket{\psi_0})[D-D_A-D_B+1]}{D-1} -\frac{1}{D_A}\,,
\end{equation}
implying the decoupling inequality:
\begin{equation}
\label{eq:decoupling_inequality_QAE}
\mathbb{E}_{\rm QAE}\left |\left | \rho_A - \frac{\mathds{1}}{D_A}  \right | \right |_1^2 \le \frac{(D_A-1)^2 + D_A(D-D_A -D_B +1)I_2(\ket{\psi_0})}{D-1}\,.
\end{equation}
Keeping $L_A/L$, $L_B/L$ fixed in the large-$L$ limit, the right-hand side of the above inequality reads
\begin{equation}
\frac{(D_A-1)^2 + D_A(D-D_A -D_B +1)I_2(\ket{\psi_0})}{D-1} \simeq \frac{D_A}{D_B} + D_A I_2(\ket{\psi_0}) = 2^{L_A -L_B} + 2^{L_A -S_2^{\rm PE}(\ket{\psi_0})}\,,
\end{equation}
where, as usual, the symbol $\simeq$ means that the two terms differ by a quantity exponentially small in $L$. Therefore, on average, the trace distance between $\rho_A$ and the completely mixed state in $\mathcal{H}$ is exponentially small as long as:
\begin{equation}
\label{eq:decoupling_RPP_condition_L_A}
L_A < {\rm min}\{L/2,\, S_2^{\rm PE}(\ket{\psi_0})\}\,.
\end{equation}
We remark that the decoupling inequality is stated using the trace distance, rather than the HS distance, as only the former has a clean interpretation in terms of distinguishability of quantum states \cite{nielsen2010quantum, watrous2018theory}. 

The decoupling inequality for the QAE, Eq. \eqref{eq:decoupling_inequality_QAE}, explains the Page curve for the average R\'enyi-2 entropy in this ensemble (as well as the 2-local circuit Page curve) found in \cite{szasz2026entanglement} (cfr. Figure 1.b of the Supplemental material): for subsystems smaller than half the total system,
 when $L_A > S_2^{\rm PE}(\ket{\psi_0})$, the average R\'enyi-2 entropy saturates to the constant value $S_2^{\rm PE}(\ket{\psi_0})$. This allows us to interpret the quantity $K:=2^{S_2^{\rm PE}(\ket{\psi_0})}$ as the effective support of the state $\ket{\psi_0}$, which coincides with the actual support if and only if $\ket{\psi_0}$ has uniform weights over $K$ basis states. Naively, the condition  \eqref{eq:decoupling_RPP_condition_L_A} means that when the effective support $K < \log_2(L/2)$, the state $\rho_A$ can be, on average, exponentially close in trace distance to the maximally mixed state in $\mathcal{H}_A$ only as long as $D_A < K$.  
 
 Since the fully mixed state is $U(1)$-invariant, the decoupling inequality also explains why $\mathbb{E}[\Delta S_A^{(2)}]=0$ when \eqref{eq:decoupling_RPP_condition_L_A} holds. However, the asymmetry remains zero also at larger subsystem sizes, namely for $L_A < L_A^*$, see Eq. \eqref{eq:asymmetry_onset_QAE}. When $L_A$ is larger than the right-hand side of Eq. \eqref{eq:decoupling_RPP_condition_L_A}, the decoupling inequality does not explain the behaviour of the asymmetry. In the following Section, we argue that the onset of the asymmetry happens when the subsystem coherence starts to become relevant. This is our second main result.

\subsection{Coherence and asymmetry onset}

In order to provide a criterion for the asymmetry onset, we turn now to the study of the average coherence content of the reduced density matrix. In particular, we are interested in estimating how much $\rho_A$ departs from being diagonal in the computational basis of $\mathcal{H}_A$ $\{\ket{s}, \, s=0,\dots,D_A-1\}$. To do so, we define the fully dephasing channel with respect to that basis, and the dephased subsystem density matrix:
\begin{equation}
\mathcal{D}_A(\cdot) := \sum_s \ket{s}\bra{s}\cdot\ket{s}\bra{s}\, \quad \omega_A := \mathcal{\mathcal{D}}_A(\rho_A)\,.
\end{equation}
Intuitively, the onset of coherence in subsystem $A$ occurs when the norm of the diagonal matrix $\omega_A$ is comparable to the norm of the matrix $\rho_A -\omega_A$, consisting of all and only the off-diagonal elements in the computational basis. Thus, we want to compare the norms $||\rho_A -\omega_A||_2$ and $||\omega_A||_2$. Notice that the dephased state $\omega_A$ is obviously closer to $\rho_A$ (in HS norm) than the fully mixed state, and in particular:
\begin{equation}
 \left|\left|\rho_A -\frac{\mathds{1}_A}{D_A}\right|\right|_2^2 = ||\rho_A -\omega_A||_2^2 + \left|\left|\omega_A -\frac{\mathds{1}_A}{D_A}\right|\right|_2^2,
\end{equation}
where the strict equality follows from the fact that the two states $\rho_A -\omega_A$ and $\omega_A -\frac{\mathds{1}_A}{D_A}$ are orthogonal in the HS product, since the first has no diagonal entries and the second is diagonal\footnote{If $A$ is diagonal with respect to a basis $\{\ket{s}\}$ and $B$ has no diagonal entries in that basis, then ${\rm Tr}(A^\dagger B)=0$}. 

Based on the above intuition, we define the onset of the average subsystem coherence in the ensemble $\mathcal{E}$ as the size $L_A^*$ such that, for $L_A > L_A^*$, one has:
\begin{equation}
\label{def:coherence_onset}
\mathbb{E}||\rho_A -\omega_A||_2^2 > \mathbb{E}||\omega_A||_2^2\,,
\end{equation}
up to terms exponentially small in the total system size $L$. In principle, the coherence onset does not need to happen at the same subsystem size where we observe the asymmetry onset. Indeed, for any reduced density matrix $\rho_A$, one can write $\rho_A=\omega_A + \chi_A^{\rm in} + \chi_A^{\rm out}$, where the two $\chi$ matrices both have only off-diagonal elements in the computational basis, but the only non-zero entries of $\chi_A^{\rm in}$ are between states belonging to the same $U(1)$ charge sector, and the only non-zero entries of $\chi_A^{\rm out}$ are between states belonging to different charge sectors. A state with $\chi_A^{\rm out}=0$ but $\chi_A^{\rm in}\ne 0$ is not coherent, but it is $U(1)$-symmetric, hence the onset of coherence can never occur at larger system sizes with respect to the onset of asymmetry. In our case, because the state $\rho = U\rho_0U^\dagger$ is evolved with a random operation that does not preserve the $U(1)$ charge, and because the average is performed with a uniform distribution over all the possible permutations and phase transformation, we do not expect any ``preference" in the onset of the off-diagonal entries. In other words, we expect the onset of coherence to coincide with the onset of asymmetry, $L_A^*$. The spread of coherence under $U(1)$-preserving Haar random dynamics was recently studied in \cite{aditya2026coherence}. 

Now, in Eq. \eqref{def:coherence_onset}, we can compute:
\begin{equation}
||\omega_A||_2^2= {\rm Tr}\,\omega_A^2\,, \quad ||\rho_A -\omega_A||_2^2 = {\rm Tr} \,\rho_A^2 - {\rm Tr} \,\omega_A^2\,,
\end{equation}
where in the second equation, we used the fact that ${\rm Tr} \,\omega_A\,\rho_A = {\rm Tr} \,\rho_A\,\omega_A={\rm Tr} \,\omega_A^2$.
The purity of $\omega_A$ is expressed in the CJ representation as follows (see Appendix \ref{sec:Choi representation for random quantum automata} and \cite{aditya2026coherence}):
\begin{equation}
\label{eq:tr_omega_A_2_main_def}
{\rm Tr\,} \omega_A^2 ={}_A\bra{\mathcal{I}^0}{}_B\braket{\mathcal{I}^+|\rho \otimes \rho}\,,
\end{equation}
hence the average over an ensemble $\mathcal{E}$ of the above quantity is:
\begin{equation}
\mathbb{E}[{\rm Tr\,} \omega_A^2] = {}_A\bra{\mathcal{I}^0}{}_B\bra{\mathcal{I}^+}\mathbb{E}[U^*\otimes U\otimes U^*\otimes U]\ket{\rho_0 \otimes \rho_0}\,. 
\end{equation}
Notice that the only difference with respect to the CJ expression of the purity of $\rho_A$, Eq. \eqref{eq:average_purity_CJ_general}, is that the 2-replica state ${}_A\bra{\mathcal{I}^-}$ is replaced by ${}_A\bra{\mathcal{I}^0}$, the CJ representation of the dephasing operator in subsystem $A$. Thus, in the random quantum automaton ensemble:
\begin{equation}
\label{eq:average_tr_omega_A_squared_QAE}
\mathbb{E}_{\rm QAE}[{\rm Tr\,} \omega_A^2] = \frac{D_B -1 + (D-D_B)I_2(\ket{\psi_0})}{D-1} \simeq 2^{-L_A} + (1-2^{-L_A})I_2(\ket{\psi_0})\,,
\end{equation}
and 
\begin{equation}
\mathbb{E}_{\rm QAE}[{\rm Tr\,} \rho_A^2-{\rm Tr\,} \omega_A^2] = \frac{(D_A -1)(1-I_2(\ket{\psi_0}))}{D-1} \simeq 2^{-L_B}(1-I_2\ket{\psi_0})\,.
\end{equation}
Again, in the above expressions, the second approximate equality holds up to exponentially decreasing terms in the system size. In the case of homogeneous product states Eq. \eqref{eq:def_homogeneous_product_state} with $\theta \ne0,\,\pi/2$, for which $I_2(\ket{\psi_0})$ is an exponentially decreasing function of $L$, one has the further asymptotic approximations:
\begin{equation}
\label{eq:coherence_onset}
2^{-L_A} + (1-2^{-L_A})I_2(\ket{\psi_0}) \simeq \begin{cases}
2^{-L_A},\, &L_A < S_2^{\rm PE}(\ket{\psi_0}) \\
I_2(\ket{\psi_0}),\, &L_A > S_2^{\rm PE}(\ket{\psi_0})
\end{cases}\,, \quad 2^{-L_B}(1-I_2(\ket{\psi_0})) \simeq 2^{-L_B}=2^{L_A-L}\,.
\end{equation}
Therefore, in the large-$L$ limit, the onset condition for coherence, Eq. \eqref{def:coherence_onset}, is verified if:
\begin{equation}
\label{eq:coherence_onset_asymptotic}
\begin{cases}
L_A > \frac{L}{2},\, &{\rm if} \quad L_A < S_2^{\rm PE} \\
L_A > L-S_2^{\rm PE}, \, &{\rm if} \quad L_A > S_2^{\rm PE}
\end{cases}\,,
\end{equation}
which is precisely equivalent to the asymmetry onset condition Eq. \eqref{eq:asymmetry_onset_QAE}.

The quantity $||\rho_A -\omega_A||_2^2 = {\rm Tr} \,\rho_A^2 - {\rm Tr} \,\omega_A^2$ is strictly related to the subsystem's R\'enyi-2 relative entropy of coherence \cite{aditya2026coherence}:
\begin{equation}
\label{eq:entropy:of_coherence_def}
C_A^{(2)}(\rho_A) := S_2(\omega_A) - S_2(\rho_A) = -\log {\rm Tr}\,\omega_A^2 +\log {\rm Tr}\,\rho_A^2\,.
\end{equation}
Notice that, for the full system, since the state $\rho$ is pure, one has:
\begin{equation}
C^{(2)}(\rho) =S_2(\mathcal{D}(\rho))= S_2^{\rm PE}(\rho),
\end{equation}
that is, the total system's coherence coincides with its participation entropy (this is of course true also for higher R\'enyi entropies as well as for the von Neumann entropy of coherence). In particular, for any realization of the operations $U$ in the QAE, $C^{(2)}(\rho)$ equals the second participation entropy of the initial state, and thus it is conserved by the dynamics. 

\begin{figure}
    \centering
    \includegraphics[width=0.45\linewidth]{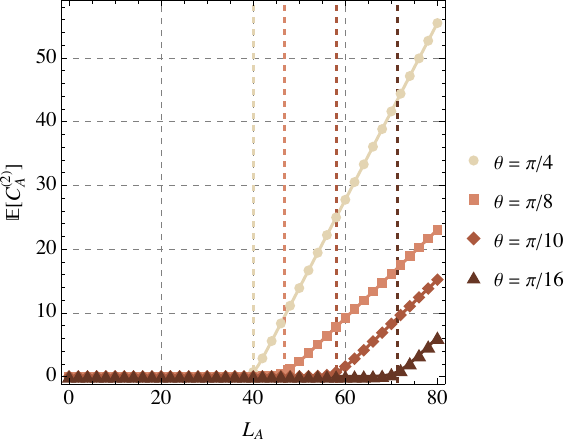}
    \caption{The coherence $C_A^{(2)}$ Page curve averaged over the 2-local quantum automaton ensemble (markers) starting from an initial product state  $\ket{\psi_0}$ \eqref{eq:def_homogeneous_product_state} for various values of the tilting angle $\theta$. The global ensemble results are shown in solid lines. The dashed lines denote the onset values $L_A^*$ as defined in Eq. \eqref{eq:asymmetry_onset_QAE}. We take $L = 80$.}
    \label{fig:coh-page}
\end{figure}

On the other hand, from the above discussion, it follows that the average subsystem's R\'enyi-2 entropy of coherence $\mathbb{E}_{\rm QAE} [C^{(2)}_A(\rho_A)]$ is zero for $L<L_A^*$ and starts to grow linearly for $L>L_A^*$, with subleading corrections around $L_A=L_A^*$, $L_A^*$ given by Eq. \eqref{eq:asymmetry_onset_QAE}. To see that this is the case, notice that Eq. \eqref{eq:average_tr_omega_A_squared_QAE} implies:
\begin{equation}
\mathbb{E}_{\rm QAE}[S_2(\omega_A)]\simeq -\log\mathbb{E}_{\rm QAE}[{\rm Tr\,} \omega_A^2] \simeq {\rm min}\{L_A,\,S_2^{\rm PE}(\ket{\psi_0})\} + O(1)\,,
\end{equation}
while the expression for the averaged purity, Eq. \eqref{eq:QAE_averaged_purity}, yields (see also \cite{szasz2026entanglement}):
\begin{equation}
\mathbb{E}_{\rm QAE}[S_2(\rho_A)]\simeq -\log\mathbb{E}_{\rm QAE}[{\rm Tr\,} \rho_A^2] \simeq {\rm min}\{L_A,\,S_2^{\rm PE}(\ket{\psi_0}),\,L-L_A\} + O(1)\,.
\end{equation}
We plot the quantity $\mathbb{E}_{\rm QAE}[ C^{(2)}_A(\rho_A)]$ for different homogeneous product states in Figure \ref{fig:coh-page}, as a function of $L_A$ and see very clearly both the distinct onset values depending on $\theta$ and the subsequent linear growth.

We conclude by observing that the dynamics of the average subsystem coherence in the 2-local QAC described in Section \ref{sec:Dynamics of entanglement asymmetry in a two-local quantum automaton circuit} can be extracted from the same equations ruling the dynamics of the average asymmetry. To be precise, one does not need the full system of ODEs \eqref{eq:G_system_short} to obtain the time evolution of the average coherence, as it is sufficient to obtain a solution for the functions $G_{n_-,n_+,n_0}(t)$ obtained by setting $n_\alpha=0$. We provide some details and numerical results in Appendix \ref{sec:Details on the dynamics in non-local QAC}.

\section{Discussion and outlook}
\label{sec:Outlook and discussions}

We have studied the average subsystem $U(1)$-asymmetry in quantum automaton ensembles with two different underlying geometries, corresponding to a global all-to-all transformation and a 2-local circuit. We have shown that the global ensemble average of the asymmetry coincides with the late-time limit of the circuit result at any system size. In a 2-local quantum automaton circuit, contrary to the random unitary case, the symmetrization of a subsystem under the dynamics occurs until subsystem's size reaches a value that depends on the initial state, namely on its participation entropy. We have derived a closed formula for this scale and quantified it for different classes of initial states. The scale at which subsystem symmetrization ceases to happen cannot be derived solely by looking at the averaged quantum distance between the reduced density matrix and the fully mixed state. Indeed, we have connected the emergence of this symmetrization scale to the growth of the average subsystem coherence and showed that they share the same characteristic length scale. Our work highlights the sharp differences between ``traditional" random unitary circuits and random quantum automata and reveals interesting new quantum phenomena.

There are several open questions raised by our work. First, it would be important to study the dynamics of entanglement asymmetry in random permutation circuits, which, on top of preserving the initial state participation entropy, display an isolated non-thermal stationary state. The tools needed to extend our results to random permutation circuits are the same applied here, although the analytical treatment of random permutations without random phases is significantly more demanding. Furthermore, since quantum automaton circuits can be implemented on current quantum platforms, investigating the growth of other relevant resources in this setting, such as nonstabilizerness and non-Gaussianity, would be extremely timely. Finally, another natural direction is to include forms of monitoring in the random automaton evolution, and see how the late time Page-curves of entanglement, asymmetry and other quantum resource monotones are affected by monitoring. 

\section{Acknowledgments}
\label{sec:acknowledgments}
MM and DSS would like to thank Filiberto Ares, Sara Murciano, Pasquale Calabrese and especially Lorenzo Piroli for fruitful discussions. This work was funded by the European Union (ERC, QUANTHEM, 101114881). Views and opinions expressed are however those of the author(s) only and do not necessarily reflect those of the European Union or the European Research Council Executive Agency. Neither the European Union nor the granting authority can be held responsible for them.

\appendix
\section{Choi-Jamio\l kowski representation for random quantum automata}
\label{sec:Choi representation for random quantum automata}

In this Appendix, we derive some results on the Choi-Jamio\l kowski representation for the random QAE which are used in the main text.
\subsection{Purities in the Choi-Jamio\l kowski representation}
Let $\mathcal{H}$ be a (finite-dimensional) Hilbert space with a basis $\{\ket{s}\}$. The identity, swap and dephasing operators in the two-replica space $\mathcal{L}(\mathcal{H}^{\otimes 2})$ are:
\begin{equation}
\mathds{1}=\sum_{s,s'}\ket{s,s'}\bra{s,s'}, \quad X=\sum_{s,s'}\ket{s,s'}\bra{s',s}, \quad \mathcal{D}=\sum_s\ket{s,s}\bra{s,s}\,.
\end{equation}
For any one-replica operator $A$, $B \in \mathcal{L}(\mathcal{H})$, the following identities hold:
\begin{equation}
{\rm Tr}_{\mathcal{H}^{\otimes 2}}[\mathds{1}(A \otimes B)]= \sum_{s,s'}\bra{s}A\ket{s}\bra{s'}B\ket{s'}={\rm Tr}_{\mathcal{H}}A\,{\rm Tr}_{\mathcal{H}}B\,,
\end{equation}
\begin{equation}
{\rm Tr}_{\mathcal{H}^{\otimes 2}}[X(A \otimes B)]= \sum_{s,s'}\bra{s}A\ket{s'}\bra{s'}B\ket{s}={\rm Tr}_{\mathcal{H}}(AB)\,,
\end{equation}
\begin{equation}
{\rm Tr}_{\mathcal{H}^{\otimes 2}}[\mathcal{D}(A \otimes B)]= \sum_s\bra{s}A\ket{s}\bra{s}B\ket{s}={\rm Tr}_{\mathcal{H}}[\mathcal{D}(A)\mathcal{D}(B)]\,,
\end{equation}
where in the last equation, we used the one-replica expression of the dephasing operator: $\mathcal{D}(A)=\sum_s \ket{s}\bra{s}A\ket{s}\bra{s}$. In particular, for two copies of the same density matrix $\rho$:
\begin{equation}
{\rm Tr}_{\mathcal{H}^{\otimes 2}}(\rho \otimes \rho) = ({\rm Tr}_\mathcal{H} \rho)^2=1, \quad {\rm Tr}_{\mathcal{H}^{\otimes 2}}[X(\rho \otimes \rho)] = {\rm Tr}_\mathcal{H} (\rho^2), \quad {\rm Tr}_{\mathcal{H}^{\otimes 2}}[\mathcal{D}(\rho \otimes \rho)] = {\rm Tr}_\mathcal{H} (\omega^2)=I_2(\rho),
\end{equation}
where $\omega:=\mathcal{D}(\rho)$ and $I_2$ is the second inverse participation ratio in the basis $\{\ket{s}\}$.

Consider now a bipartite space $\mathcal{H}=\mathcal{H}_A\otimes \mathcal{H}_B$, with a basis $\{\ket{\alpha}\otimes\ket{\beta}\equiv \ket{\alpha\,\beta}\}$. We define the two-replica operators $X_A$ and $\mathcal{D}_A$, respectively acting as the swap and dephasing operator in $\mathcal{H}_A$, and as the identity in $\mathcal{H}_B$:
\begin{equation}
X_A = \sum_{\alpha,\alpha',\beta,\beta'}\ket{\alpha\,\beta\,\alpha'\,\beta'}\bra{\alpha'\,\beta\,\alpha\,\beta'}\,, \quad \mathcal{D}_A =\sum_{\alpha,\alpha',\beta,\beta'}\ket{\alpha\,\beta\,\alpha\,\beta'}\bra{\alpha\,\beta\,\alpha\,\beta'}\,,
\end{equation}
with the convention that $\ket{\alpha\,\beta\,\delta\,\gamma}=\ket{\alpha\,\beta}\otimes\ket{\delta\,\gamma}$, where $\ket{\alpha\,\beta}$ is in the first copy of $\mathcal{H}$, and $\ket{\delta\,\gamma}$ is in the second. Thus, the purity of the reduced density matrix $\rho_A$, and the purity of the dephased reduced density matrix $\omega_A=\mathcal{D}_A(\rho_A)$ can be expressed as:
\begin{align}
\label{eq:appendix_swap_purity}
{\rm Tr}_{\mathcal{H}^{\otimes 2}}[X_A(\rho \otimes \rho)] &= \sum_{\alpha,\alpha',\beta,\beta'} \bra{\alpha'\,\beta\,\alpha\,\beta'} \rho \otimes \rho \ket{\alpha\,\beta\,\alpha'\,\beta'} = \sum_{\alpha,\alpha',\beta,\beta'} \bra{\alpha'\,\beta}\rho\ket{\alpha\,\beta}\bra{\alpha\,\beta'}\rho\ket{\alpha'\beta'} \nonumber \\
&=\sum_{\alpha,\alpha'}\bra{\alpha'}{\rm Tr}_{\mathcal{H}_B}\rho \ket{\alpha}\bra{\alpha}{\rm Tr}_{\mathcal{H}_B}\rho \ket{\alpha'}= {\rm Tr}_{\mathcal{H}_A}\rho_A^2\,,
\end{align}
\begin{align}
\label{eq:appendix_dephasing_purity}
{\rm Tr}_{\mathcal{H}^{\otimes 2}}[\mathcal{D}_A(\rho \otimes \rho)] &= \sum_{\alpha,\beta,\beta'} \bra{\alpha\,\beta\,\alpha\,\beta'} \rho \otimes \rho \ket{\alpha\,\beta\,\alpha\,\beta'} = \sum_{\alpha,\beta,\beta'} \bra{\alpha\,\beta}\rho\ket{\alpha\,\beta}\bra{\alpha\,\beta'}\rho\ket{\alpha\beta'} \nonumber \\
&=\sum_{\alpha}\bra{\alpha}{\rm Tr}_{\mathcal{H}_B}\rho \ket{\alpha}\bra{\alpha}{\rm Tr}_{\mathcal{H}_B}\rho \ket{\alpha}= {\rm Tr}_{\mathcal{H}_A}\omega_A^2\,.
\end{align}

The Choi-Jamio\l kowski isomorphism \cite{watrous2018theory} is a linear map that associates a state to any completely positive operator. In particular, in one replica, it is a map from $\mathcal{L}(\mathcal{H})$ to $\mathcal{H}^{\otimes 2}$, and the density matrix operator $\rho$ is mapped to:
\begin{equation}
\ket{\rho} := \mathds{1}\otimes \rho \ket{\mathcal{I}^+}\,, \quad \ket{\mathcal{I}^+}=\sum_s \ket{s}\otimes\ket{s}\,,
\end{equation}
that is, $\ket{\mathcal{I}^+}$ is the un-normalized maximally entangled state in $\mathcal{H}^{\otimes 2}$, which represents the identity. If $\mathcal{H}$ is the tensor product of local Hilbert spaces $\mathcal{H}_k$, then $\ket{\mathcal{I}^+}=\otimes_k\ket{I^+}_k$. It is straightforward to check that in this formalism ${\rm Tr}(\sigma^\dagger\rho)=\braket{\sigma|\rho}$, and in particular ${\rm Tr}\rho = \braket{\mathcal{I}^+|\rho}$. 
The isomorphism is easily extended to the case where there are more replicas of the original Hilbert space. in particular, for two replicas, the isomorphism becomes a map from $\mathcal{L}(\mathcal{H}^{\otimes 2})$ to $\mathcal{H}^{\otimes 4}$ and:
\begin{equation}
\ket{\rho \,\otimes \rho} = \mathds{1}\otimes \rho \otimes \mathds{1} \otimes \rho \ket{\mathcal{I}^+}
\end{equation}
where we adopt the same notation $\ket{\mathcal{I}^+}$ to represent the tensor product of two fully entangled states in two copies of $\mathcal{H}^{\otimes 2}$:
\begin{equation}
\ket{\mathcal{I}^+} = \sum_{s,s'}\ket{s\,s\,s'\,s'}\,.
\end{equation}
In general, it will be clear from the context whether  $\ket{\mathcal{I}^+}$ is a state in $\mathcal{H}^{\otimes 2}$ or in $\mathcal{H}^{\otimes 4}$. The swap and the dephasing operators $X,\,\mathcal{D} \in \mathcal{L}(\mathcal{H}^{\otimes 2})$ are mapped to:
\begin{equation}
\ket{\mathcal{I}^-}=\sum_{s,s'}\ket{s\,s'\,s'\,s}\,, \quad \ket{\mathcal{I}^0}=\sum_s\ket{s\,s\,s\,s}\,,
\end{equation}
as one can immediately check that:
\begin{equation}
{\rm Tr}_{\mathcal{H}^{\otimes 2}}[X(\rho \otimes \rho)] = \braket{\mathcal{I}^-|\rho \otimes \rho}\,, \quad {\rm Tr}_{\mathcal{H}^{\otimes 2}}[\mathcal{D}(\rho \otimes \rho)] = \braket{\mathcal{I}^0|\rho \otimes \rho}\,.
\end{equation}
If there is a spatial bipartition $\mathcal{H}=\mathcal{H}_A\otimes \mathcal{H}_B$ then, defining $\ket{\mathcal{I}^+}_{A/B}=\otimes_{k \in A/B} \ket{I^+}_k$, it is also easy to check:
\begin{equation}
{\rm Tr}_{\mathcal{H}^{\otimes 2}}[X_A(\rho \otimes \rho)] = {}_A\bra{\mathcal{I}^-}{}_B\braket{\mathcal{I}^+|\rho \otimes \rho}\,, \quad {\rm Tr}_{\mathcal{H}^{\otimes 2}}[\mathcal{D}_A(\rho \otimes \rho)] = {}_A\bra{\mathcal{I}^0}{}_B\braket{\mathcal{I}^+|\rho \otimes \rho}\,.
\end{equation}
Combining the above equations with Eqs. \eqref{eq:appendix_swap_purity} and \eqref{eq:appendix_dephasing_purity}, we obtain the expression of the purities ${\rm Tr}_{\mathcal{H}_A}\rho_A^2$ and ${\rm Tr}_{\mathcal{H}_A}\omega_A^2$ in the Choi-Jamio\l kowski representation:
\begin{equation}
{\rm Tr}_{\mathcal{H}_A}\rho_A^2 = {}_A\bra{\mathcal{I}^-}{}_B\braket{\mathcal{I}^+|\rho \otimes \rho}\,, \quad {\rm Tr}_{\mathcal{H}_A}\omega_A^2 = {}_A\bra{\mathcal{I}^0}{}_B\braket{\mathcal{I}^+|\rho \otimes \rho}\,.
\end{equation}

\subsection{First and second moments}
In this appendix, we compute the expectation values $\mathbb{E}[U^*\otimes U]$ and $\mathbb{E}[U^*\otimes U\otimes U^*\otimes U]$, where $U,\,U^* \in U(D)$ are drawn from the random quantum automaton ensemble $\mathcal{E}_{\rm QAE}$. Here, $U^*$ denotes the complex conjugate of $U$. First, let us recall the useful identity for the Choi-Jamio\l kowski representation of a product: 
\begin{equation}
\ket{ABC}=C^{\rm T}\otimes A\ket{B}\,,
\end{equation}
which holds for every completely positive operator $A$, $B$, $C\in \mathcal{L}(\mathcal{H})$. It is easy to see that the above is true, as for $\ket{i,j}\in \mathcal{H}^{\otimes 2}$ on one hand:
\begin{equation}
\bra{i,j}\mathds{1}\otimes ABC \ket{\mathcal{I}^+} = \sum_s\braket{i|s}\braket{j|ABC|s}=\braket{j|ABC|i}\,,
\end{equation}
and on the other hand:
\begin{equation}
\bra{i,j}C^{\rm T}\otimes A\ket{B}=\bra{i,j}C^{\rm T}\otimes AB\ket{\mathcal{I}^+}=\sum_s \bra{i}C^{\rm T}\ket{s}\bra{j}AB\ket{s}=\sum_s\bra{s}C\ket{i}\bra{j}AB\ket{s}=\bra{j}ABC\ket{i}\,.
\end{equation}
Thus, in particular, for a state $\rho=U\rho_0 U^\dagger$, in one and two replicas the following holds:
\begin{equation}
\ket{U\rho U^\dagger}=U^*\otimes U \ket{\rho_0}, \quad \ket{U\rho U^\dagger \otimes U\rho U^\dagger}=U^*\otimes U \otimes U^*\otimes U \ket{\rho_0\otimes \rho_0}. 
\end{equation}
Let us then compute the moments:
\begin{equation}
\mathcal{M}_1 := \mathbb{E}_{\rm QAE}[U^*\otimes U]\,, \quad \mathcal{M}_2 := \mathbb{E}_{\rm QAE}[U^*\otimes U \otimes U^*\otimes U]\,,
\end{equation}
where, concretely, the average is over the set of random phases and random permutations of the computational basis states. 

Let us consider $\mathcal{M}_1$. On a computational basis state $\ket{a}$, $a=0,\dots,D-1$:
\begin{equation}
U\ket{a} = e^{i\phi_a}\ket{\pi(a)}\,, \quad U^*\ket{a} = e^{-i\phi_a}\ket{\pi(a)}\,,
\end{equation}
for some permutation $\pi \in \mathcal{S}_D$ and some $\phi_a \in [-\pi,\pi)$, thus on a basis state $\ket{a\,b} \in \mathcal{H}^{\otimes 2}$:
\begin{equation}
\mathcal{M}_1\ket{a\,b} = \mathbb{E}_{\{\phi_a,\phi_b\}}\mathbb{E}_\pi e^{-i\phi_a+i\phi_b}\ket{\pi(a)\pi(b)}\,.
\end{equation}
The average over the phases selects $a=b$. In fact, if $a\ne b$ then the phases are independent and $\mathbb{E}[e^{-i\phi_a+i\phi_b}]=\mathbb{E}e^{-i\phi_a}\mathbb{E}e^{i\phi_b}=\int_{-\pi}^\pi d\phi_a e^{-i\phi_a}\int_{-\pi}^\pi d\phi_b e^{i\phi_b}=0$. Then, the average over permutations is the sum over $\pi$ of $\ket{\pi(a)\pi(a)}$, divided by the total number of permutations, which is  $D!$. For every label $a'=0,\dots,D-1$, there are exactly $(D-1)!$ permutations $\pi$ that keep $a'=\pi(a)$ fixed. Thus:
\begin{equation}
\mathcal{M}_1\ket{a\,b} = \delta_{a,b} (D!)^{-1}\sum_\pi\ket{\pi(a)\pi(a)}=\delta_{a,b} (D!)^{-1}\sum_{a'}(D-1)!\ket{a'\,a'}=\frac{\delta_{a,b}}{D}\ket{\mathcal{I}^+}\,,
\end{equation}
yielding:
\begin{equation}
\label{eq:first_moment_QAE}
\mathcal{M}_1 = \frac{\ket{\mathcal{I}^+}\bra{\mathcal{I}^+}}{D}\,,
\end{equation}
which in operatorial form is the fully mixed state $\mathds{1}_{\mathcal{H}}/D$.

We now compute $\mathcal{M}_2$. On a basis state $\ket{a\,b\,c\,d}\in\mathcal{H}^{\otimes 4}$, the action of $\mathcal{M}_2$ reads:
\begin{equation}
\mathcal{M}_2\ket{a\,b\,c\,d} = \mathbb{E}_{\{\phi\}}\mathbb{E}_\pi e^{-i\phi_a+i\phi_b-i\phi_c+i\phi_d}\ket{\pi(a)\pi(b)\pi(c)\pi(d)}\,.
\end{equation}
The average over the phases yields a nonzero contribution if the sets $\{a,c\}=\{b,d\}$, which happens in three distinct cases: $a=b=c=d$, or $a=b \ne c=d$, or $a=d\ne b=c$. Consider first the case $a=b=c=d$. This computation is completely analogous to that of $\mathcal{M}_1$:
\begin{equation}
\mathcal{M}_2\ket{a\,a\,a\,a} = (D!)^{-1}\sum_\pi\ket{\pi(a)\pi(a)\pi(a)\pi(a)}= (D!)^{-1}\sum_{a'}(D-1)!\ket{a'\,a'\,a'\,a'}=\frac{\ket{\mathcal{I}^0}}{D}\,.
\end{equation}
Next, consider the case $a=b \ne c=d$, where $\mathcal{M}_2\ket{a\,a\,c\,c}=\mathbb{E}_\pi \ket{\pi(a)\pi(a)\pi(c)\pi(c)}$. Because the permutation $\pi$ is a bijection, if $a\ne c$ then $\pi(a)\ne\pi(c)$, and there are $(D-2)!$ out of the $D!$ total permutations that map the pair $(a,c)$ to $(a',c')$, with $a'\ne c'$. Thus, for $a\ne c$:
\begin{equation}
\mathcal{M}_2\ket{a\,a\,c\,c} = \frac{1}{D!}\sum_\pi \ket{\pi(a)\pi(a)\pi(c)\pi(c)}=\frac{1}{D!}\sum_{a'\ne c'} (D-2)!\ket{a'\,a'\,c'\,c'} = \frac{\ket{\mathcal{I}^+}-\ket{\mathcal{I}^0}}{D(D-1)}\,.
\end{equation}
The case $a=d\ne b=c$ is analogous to the one above, leading to (for $a\ne b$):
\begin{equation}
\mathcal{M}_2\ket{a\,b\,b\,a} = \frac{1}{D!}\sum_{a'\ne b'} (D-2)!\ket{a'\,b'\,b'\,a'} = \frac{\ket{\mathcal{I}^-}-\ket{\mathcal{I}^0}}{D(D-1)}\,.
\end{equation}
Putting everything together:
\begin{align}
\label{eq:second_moment_QAE}
\mathcal{M}_2 &= \frac{\ket{\mathcal{I}^0}\bra{\mathcal{I}^0}}{D}+\frac{(\ket{\mathcal{I}^+}-\ket{\mathcal{I}^0})(\bra{\mathcal{I}^+}-\bra{\mathcal{I}^0})}{D(D-1)}+\frac{(\ket{\mathcal{I}^-}-\ket{\mathcal{I}^0})(\bra{\mathcal{I}^-}-\bra{\mathcal{I}^0})}{D(D-1)} \nonumber \\
&= \frac{\ket{\mathcal{I}^+}\bra{\mathcal{I}^+} + \ket{\mathcal{I}^-}\bra{\mathcal{I}^-} + (D+1)\ket{\mathcal{I}^0}\bra{\mathcal{I}^0}-(\ket{\mathcal{I}^-}\bra{\mathcal{I}^0}+\ket{\mathcal{I}^0}\bra{\mathcal{I}^-}+\ket{\mathcal{I}^+}\bra{\mathcal{I}^0}+\ket{\mathcal{I}^0}\bra{\mathcal{I}^+})}{D(D-1)}\,,
\end{align}
which is exactly Eq. \eqref{eq:second_moment_QAE_main} in the main text.

We mention that the above expectation values, Eqs. \eqref{eq:first_moment_QAE} and \eqref{eq:second_moment_QAE}, can also be deduced from the Weingarten matrix of the random permutation ensemble \cite{Bertini2025Permutation} by selecting, among the invariant 2-replica states in that ensemble, those which are also invariant under random phase transformations.

\section{Details on the dynamics in non-local QAC}
\label{sec:Details on the dynamics in non-local QAC}
In this Section, we provide the details of the derivation of Eq. \eqref{eq:G_system_short}, which closely follows the derivation of the analogous system for the averaged subsystem purities in the Supplemental material of \cite{szasz2026entanglement}.
\subsection{Derivation of the differential equations for the charged moments}
 We start by explicitly writing the action of the permutation operator $\hat{\pi}$ in Eq. \eqref{eq:G_state_definition} as follows:
\begin{equation}
\ket{G_{n_\alpha, n_-, n_+, n_0}^\alpha}= \frac{1}{L!} \sum_{\pi \in {\mathcal{S}}_L} \bigotimes_{j \in A_\pi^{\alpha}} \ket{I_\alpha^-}_j \bigotimes_{j \in A_\pi^-} \ket{I^{-}}_j \bigotimes_{j \in A_\pi^+} \ket{I^+}_j \bigotimes_{j \in A_\pi^0} \ket{I^0}_j\,,
\label{state}
\end{equation} 
where $A_\pi^\alpha =\{\pi(1),\dots,\pi(n_\alpha)\}$, $A_\pi^- =\{\pi(n_\alpha +1),\dots,\pi(n_\alpha +n_-)\}$, $A_\pi^+ =\{\pi(n_\alpha +n_-+1),\dots,\pi(n_\alpha +n_-+n_+)\}$, $A_\pi^0 =\{\pi(n_\alpha +n_-+n_++1),\dots,\pi(L)\}$. In order to evaluate the action of $\sum_{1\le j < k\le L} \mathcal{U}_{j,k}$ on $\bra{G_{n_\alpha, n_-, n_+, n_0}^\alpha}$, for each permutation $\pi$ we split the sum over $j$ and $k$ in the following way:
\begin{align}
 \sum_{1\leq j<k\leq L} \langle G_{n_\alpha, n_-, n_+, n_0}^\alpha|\mathcal{U}_{j,k} &= \frac{1}{L!} \sum_{\pi \in {\mathcal{S}}_N} \left[\sum_{\substack{j,k\in A_\pi^\alpha \\ j < k}}+\sum_{\substack{j,k\in A_\pi^- \\ j < k}}+ \sum_{\substack{j,k\in A_\pi^+ \\ j < k}}+\sum_{\substack{j,k\in A_\pi^0 \\ j < k}}  + \sum_{\substack{j \in A_\pi^\alpha \\ k \in A_\pi^-}} +\sum_{\substack{j \in A_\pi^\alpha \\ k \in A_\pi^+}}+\sum_{\substack{j \in A_\pi^\alpha \\ k \in A_\pi^0}} + \sum_{\substack{j \in A_\pi^- \\ k \in A_\pi^+}} +\sum_{\substack{j \in A_\pi^- \\ k \in A_\pi^0}}+\sum_{\substack{j \in A_\pi^+ \\ k \in A_\pi^0}} \right]
 \nonumber\\
& \bigotimes_{j \in A_\pi^\alpha} {}_j\bra{I_\alpha^-}
\bigotimes_{j \in A_\pi^-} {}_j\bra{I^-}\bigotimes_{j \in A_\pi^+} {}_j\bra{I^+} \bigotimes_{j \in A_\pi^0} {}_j\bra{I^0}\,\mathcal{U}_{j,k}\,.
\end{align} 
We can now proceed exactly as in \cite{szasz2026entanglement}, using the relations \eqref{eq:QAC_overlaps_old} and \eqref{eq:QAC_overlaps_alpha}:  
\begin{align}
 &\sum_{1\leq j<k\leq L} \langle G_{n_\alpha, n_-, n_+, n_0} ^\alpha|\mathcal{U}_{j,k} = \left(\frac{n_-(n_- -1)}{2}+\frac{n_+(n_+-1)}{2}+\frac{n_0(n_0-1)}{2} \right) \langle G_{n_\alpha, n_-, n_+, n_0}^\alpha|\nonumber\\
& + \frac{n_\alpha(n_\alpha-1)}{2}\left(A(\alpha)  \langle G_{n_\alpha-2, n_- + 2, n_+, n_0}^\alpha| + B(\alpha) \langle G_{n_\alpha-2, n_- , n_+, n_0+2}^\alpha|\right) \nonumber\\
&+ \frac{n_\alpha n_-}{3}\left(F(\alpha) \langle  G_{n_\alpha-1,n_-+1, n_+ , n_0}^\alpha|+ G(\alpha)  \, \langle G_{n_\alpha-1, n_- -1, n_+, n_0+2}^\alpha|\right) \nonumber \\
& + \frac{n_\alpha n_+}{3}\left(\langle G_{n_\alpha-1, n_-, n_+ +1, n_0}^\alpha|+C(\alpha) \, \langle G_{n_\alpha-1, n_-+2, n_+-1, n_0}^\alpha|+D(\alpha) \, \langle G_{n_\alpha-1,n_-, n_+-1, n_0+2}^\alpha|\right)\nonumber\\
& 
+ \frac{n_\alpha n_0}{3}\left(C(\alpha)\, \langle G_{n_\alpha-1,n_-+2, n_+ , n_0-1}^\alpha|+q E(\alpha) \, \langle G_{n_\alpha-1, n_-, n_+, n_0+1}^\alpha|\right)\nonumber\\
& 
+ \frac{n_- n_+}{3}\left(\langle G_{n_\alpha, n_--1, n_+ +1, n_0}^\alpha|+ \langle G_{n_\alpha, n_-+1, n_+-1, n_0}^\alpha|+ \, \langle G_{n_\alpha, n_--1, n_+-1, n_0+2}^\alpha|\right)\nonumber\\
&
+ \frac{n_- n_0}{3}\left(\langle G_{n_\alpha, n_-+1, n_+ , n_0-1}^\alpha|+2 \, \langle G_{n_\alpha, n_--1, n_+, n_0+1}^\alpha|\right) \nonumber \\
& + \frac{n_+ n_0}{3}\left(\langle G_{n_\alpha,n_-, n_+ +1, n_0-1}^\alpha|+2 \, \langle G_{n_\alpha, n_-, n_+-1, n_0+1}^\alpha|\right)\,.
\end{align} 
 Adding to the above the contribution coming from the identity in Eq. \eqref{eq:Lindbladian} and taking the overlap with $\ket{\rho(t)\otimes \rho(t)}$, we end up with the following differential equation:
\begin{align}
\frac{d G_{n_\alpha, n_-, n_+, n_0}^\alpha(t)}{dt} &=\frac{n_\alpha +n_-^2 + n_+^2+n_0^2-L^2}{L-1} G_{n_\alpha, n_-, n_+, n_0}^\alpha(t) \nonumber\\
& + \frac{n_\alpha(n_\alpha-1)}{L-1}\left(A(\alpha) G_{n_\alpha-2, n_- + 2, n_+, n_0}^\alpha(t) + B(\alpha) G_{n_\alpha-2, n_- , n_+, n_0+2}^\alpha(t)\right) \nonumber\\
&
+ \frac{2 n_\alpha n_-}{3(L-1)}\left(F(\alpha)  G_{n_\alpha-1,n_-+1, n_+ , n_0}^\alpha(t)+
G(\alpha)  \, G_{n_\alpha-1, n_- -1, n_+, n_0+2}^\alpha(t)\right) \nonumber \\
& + \frac{2 n_\alpha n_+}{3(L-1)}\left(G_{n_\alpha-1, n_-, n_+ +1, n_0}^\alpha(t)+C(\alpha) \, G_{n_\alpha-1, n_-+2, n_+-1, n_0}^\alpha(t) +D(\alpha) \, G_{n_\alpha-1,n_-, n_+-1, n_0+2}^\alpha(t)\right)\nonumber\\
&+ \frac{2 n_\alpha n_0}{3(L-1)}\left(C(\alpha)\, G_{n_\alpha-1,n_-+2, n_+ , n_0-1}^\alpha(t)+E(\alpha) \,  G_{n_\alpha-1, n_-, n_+, n_0+1}^\alpha(t)\right)\nonumber\\
& 
+ \frac{2 n_- n_+}{3(L-1)}\left(G_{n_\alpha, n_--1, n_+ +1, n_0}^\alpha(t)+  G_{n_\alpha, n_-+1, n_+-1, n_0}^\alpha(t)+ \, G_{n_\alpha, n_--1, n_+-1, n_0+2}^\alpha(t)\right)\nonumber\\
& 
+ \frac{2 n_- n_0}{3(L-1)}\left( G_{n_\alpha, n_-+1, n_+ , n_0-1}^\alpha(t)+2 \,  G_{n_\alpha, n_--1, n_+, n_0+1}^\alpha(t)\right) \nonumber \\
& + \frac{2 n_+ n_0}{3(L-1)}\left(G_{n_\alpha,n_-, n_+ +1, n_0-1}^\alpha(t)+q \, G_{n_\alpha, n_-, n_+-1, n_0+1}^\alpha(t)\right)\,,
\label{eq:G_system_long}
\end{align} 
which reproduces Eq. \eqref{eq:G_system_short}.
\subsection{Dynamics of the averaged purity and R\'enyi-2 entropy of coherence}
As mentioned in Section \ref{sec:Dynamics of entanglement asymmetry in a two-local quantum automaton circuit}, the dynamics of the average purities of all subsystems can be obtained by solving a simpler, closed set of equations for the functions
$G_{n_-,n_+,n_0}(t) := G^{\alpha=0}_{n_\alpha=0,n_-,n_+,n_0}(t)$. This system was derived in \cite{szasz2026entanglement}, and we report it below:
\begin{align}
\frac{d G_{n_-, n_+, n_0}(t)}{dt} &=\frac{ n_-^2 + n_+^2+n_0^2-L^2}{L-1} G_{n_-, n_+, n_0}(t) \nonumber\\
& 
+ \frac{2 n_- n_+}{3(L-1)}\left(G_{n_--1, n_+ +1, n_0}(t)+  G_{n_-+1, n_+-1, n_0}(t)+ \, G_{n_--1, n_+-1, n_0+2}(t)\right)\nonumber\\
& 
+ \frac{2 n_- n_0}{3(L-1)}\left( G_{n_-+1, n_+ , n_0-1}(t)+2 \,  G_{n_--1, n_+, n_0+1}(t)\right) \nonumber \\
& + \frac{2 n_+ n_0}{3(L-1)}\left(G_{n_-, n_+ +1, n_0-1}(t)+q \, G_{n_-, n_+-1, n_0+1}(t)\right)\,.
\label{eq:G_system_purities}
\end{align} 
For a permutation invariant initial state, the average purity at time $t$, $\mathbb{E}_{\rm QAC} {\rm Tr} \rho_A^2(t)$ is the same for all subsystems $A$ such that $|A|=L_A$ and it is given by $\mathbb{E}_{\rm QAC} {\rm Tr} \rho_A^2(t)= G_{L_A,L-L_A,0}(t)$.

\begin{figure}
    \centering
    \includegraphics[width=1.\linewidth]{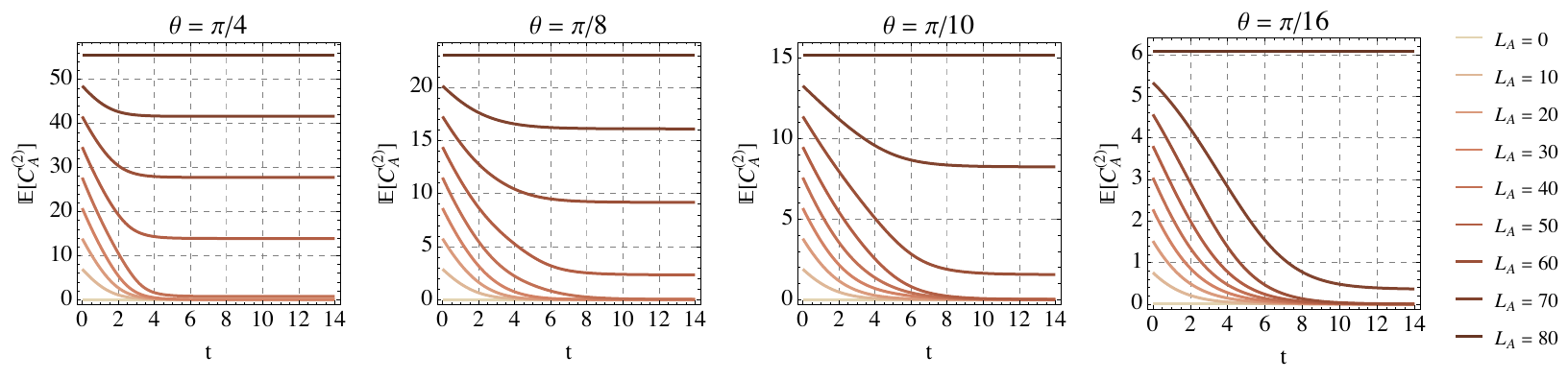}
   \caption{The averaged coherence $C_A^{(2)}$ under 2-local random quantum automaton circuit evolution starting from an initial product state $\ket{\psi_0}$ \eqref{eq:def_homogeneous_product_state} for various values of the tilting angle $\theta$. We take $L = 80$.}
    \label{fig:coh-dynamics}
\end{figure}

Remarkably, once the system Eq. \eqref{eq:G_system_purities} is solved, as a byproduct one also obtains all the average R\'enyi-2 entropies of coherence. To be precise, from the definition Eq. \eqref{eq:tr_omega_A_2_main_def} and using permutation invariance, we find that:
\begin{equation}
\mathbb{E}_{\rm QAC} {\rm Tr} \omega_A^2(t) = G_{0,L-L_A,L_A}(t)\,,
\end{equation}
and in particular $\forall\,t$, $G_{0,0,L}(t)=I_2(\ket{\psi_0})$. The time-profile of the average R\'enyi-2 entropy of coherence, see Eq. \eqref{eq:entropy:of_coherence_def}, can be then obtained using the self-averaging properties of the QAC ensemble:
\begin{equation}
\mathbb{E}_{\rm QAC}[C_A^{(2)}(t)] \simeq   -\log \mathbb{E}_{\rm QAC} [{\rm Tr\,} \omega_A^2(t)] +\log \mathbb{E}_{\rm QAC} [{\rm Tr\,} \rho_A^2(t)]\,.
\end{equation}
The numerical results for the time-evolution of the average subsystem coherence, starting from an homogeneous product state, are shown in Figure \ref{fig:coh-dynamics}. 

\bibliography{bibliography}
\end{document}